\documentclass[preprint,number,sort&compress]{elsarticle}
\usepackage{graphicx}
\usepackage{amsmath}
\usepackage{color}
\usepackage{cprotect}
\usepackage{multirow}
\usepackage{enumerate}
\usepackage{natbib}
\usepackage[pagewise]{lineno}

\journal{Mathematical Biosciences}

\newcommand{\av}[1]{\left\langle {#1 }\right\rangle}  

\newcommand{\e}{\varepsilon}              %
\newcommand{\w}{\omega}              %
\newcommand{\G}{\Gamma}              %
\renewcommand{\l}{\lambda}

\begin{document}

\title{Statistical Theory of Asymmetric Damage Segregation in Clonal Cell Populations}


\author[1]{Arkady Pikovsky}
\address[1]{Department of Physics and Astronomy, University of Potsdam, Karl-Liebknecht-Strasse 24/25, 14476, Potsdam-Golm, Germany}
\ead{pikovsky@uni-potsdam.de}
\author[2]{Lev S. Tsimring\corref{cor1}}
\ead{ltsimring@ucsd.edu }
\address[2]{BioCircuits Institute, University of California San Diego, 9500 Gilman Drive, La Jolla, CA 92093-0328, USA}
\cortext[cor1]{Corresponding author}
\date{\today}
\begin{abstract}
Asymmetric damage segregation (ADS) is ubiquitous among unicellular organisms: After a mother cell divides, its two daughter cells receive sometimes slightly, sometimes strongly different fractions of damaged proteins accumulated in the mother cell. Previous studies demonstrated that ADS provides a selective advantage over symmetrically dividing cells by rejuvenating and perpetuating the population as a whole. In this work we focus on the statistical properties of damage in individual lineages and the overall damage distributions in growing populations for a variety of ADS models with different rules governing damage accumulation, segregation, and the lifetime dependence on damage. We show that for a large class of deterministic ADS rules the trajectories of damage along the lineages are chaotic, and the distributions of damage in cells born at a given time asymptotically becomes fractal. By exploiting the analogy of linear ADS models with the Iterated Function Systems known in chaos theory, we derive the Frobenius-Perron equation for the stationary damage density distribution and analytically compute the damage distribution moments and fractal dimensions. We also investigate nonlinear and stochastic variants of ADS models and show the robustness of the salient features of the damage distributions. 
\end{abstract}

\maketitle

\section{Introduction}
Many unicellular organisms among bacteria and yeasts proliferate by binary fission so that
each mother cells divides into two seemingly identical daughter cells. However, more careful examination of cell  
lineages \cite{stewart2005aging,ackermann2007evolutionary} revealed that the progeny is in fact slightly asymmetric, one daughter cell has a 
slightly longer lifespan than the other. This difference, at least in rode-shaped bacteria {\em E. coli},  
is apparently rooted in the fact that every bacterium is itself slightly asymmetric since its two poles  always 
have a different age: the old pole that existed in its mother cell and the new pole that was  created when its 
mother cell divided. The ``age" of a cell can be defined as the age of the old pole (the number of generations 
it has been in existence). It was found that the daughter cell that inherits the 
older pole from its mother, grows slower and divides slightly later than the daughter cell that  inherits 
the newer pole. A plausible hypothesis that can explain such a correlation is that the cells  gradually 
accumulate damaged proteins which aggregates near their poles, and therefore the daughter cell that inherits 
the older pole, will also inherit a larger fraction of damaged proteins hitherto accumulated in the mother cell.  
Direct visualization of protein aggregates in growing cell lineages  corroborates this conjecture \cite{aguilaniu2003asymmetric,lindner2008asymmetric,winkler2010quantitative,coelho2014fusion}. 

A number of computational  and theoretical studies addressed the dynamics of asymmetric  damage segregation (ADS) in a growing microbial population and their implications for the overall population fitness 
\cite{evans2007damage,erjavec2008selective,chao2010model,strandkvist2014asymmetric,vedel2016asymmetric,lin2019optimal,blitvic2020aging,min2021transport,levien2021non}.  
In \cite{chao2010model,vedel2016asymmetric} it was demonstrated numerically that asymmetry accelerates the average growth of the population as a whole. In \cite{min2021transport}, a kinetic description of damage segregation was developed on the basis of a transport equation for the time-dependent damage distribution function that was applied both to deterministic and stochastic damage synthesis and segregation. Using that equation, moments and the average population growth rate were computed analytically in the limit of small asymmetry. However, a comprehensive understanding of the structure of damage distribution in a population of asymmetrically dividing cells is still lacking. In this paper, we focus on deterministic asymmetric division and analyze this structure using a general Frobenius-Perron equation for the at-birth damage distribution function. It turns out that if the rules controlling damage accumulation and inheritance are deterministic, the system governing the damage distribution is analogous, and for certain class of linear damage accumulation and division rules exactly equivalent, to the Iterated Function System (IFS) known in the theory of fractals~\cite{barnsley2014fractals}. Exploiting this analogy, we show that the asymptotic stationary distribution of damage is indeed fractal and find the spectrum of its fractal dimensions. For more general nonlinear models of damage accumulations and segregation we analyze the structure of the damage distribution functions numerically and show that the its fractal nature is robust. We discuss also deterministic irregularity of the damages  in a lineage and relate it, for the linear damage redistribution rule, to chaotic properties of the IFS. If the damage accumulation and/or segregation have stochastic components, the distribution smears out but remains multi-peaked.     

\section{Deterministic Models of ADS}
\label{sec:models}
In this section we introduce a class of models for the asymmetric damage segregation 
that will form the basis of our theory.  We will suppose that a 
cell is created at time $t_0$ and divides at time $t_0+\tau$. We denote the 
instantaneous cell damage (it can be, for example, the amount of damaged proteins, which is assumed to be a real number) $D(t)$, where $t_0\leq t \leq t_0+\tau$.  The initial cell damage (inherited from the mother cell) 
is $x=D(t_0)$, and the final damage just before division is $y=D(t_0+\tau)$, 
where $\tau$ is the cell lifetime (the latter is constant in some setups or damage-dependent in other formulations). 

There are three components in each model that determines the distribution of damage within the population of growing 
and dividing cells:  
\begin{enumerate}[{\bf 1.}]
\item \textbf{Damage gain} specifies temporal dynamics of damage $D(t;x)$ in a cell starting from its value $x$ at
birth at $t=t_0$ to division at $t=t_0+\tau$. We will denote the final cell damage as $y=D(t_0+\tau;x)$. 

{\em Examples:}
\begin{enumerate}[a.]
\item Vedel et. al model \cite{vedel2016asymmetric}. In this model it it assumed that 
every cell adds a fixed amount of damage $\lambda$ over its lifetime,  $y=x+\lambda$. The damage is added with the constant rate $\gamma$, so that for a lifetime  $\tau$ this rate is $\gamma=\lambda/\tau$. The evolution of the damage is thus $D(t;x)=x+\lambda (t-t_0)/\tau$.
\item Damage obeys a linear differential equation $\dot D=\beta D+\gamma$, where the term $\beta D$ 
describes auto-catalytic production or degradation of damage (depending on the sign of $\beta$), with the initial condition $D(t_0)=x$.
\end{enumerate}
\item \textbf{Lifetime} $\tau$ might be constant (damage-independent), or it can be a functional of the damage $D(t)$.
Since in deterministic models at any moment of time the 
current damage $D(t;x)$ is pre-determined by the initial value $x$ at the instant 
of birth $t_0$, we can assume that the lifetime is a function of the initial damage, $\tau(x)$.

{\em Examples:} 
\begin{enumerate}[a.]
\item  In model~\cite{vedel2016asymmetric}, the lifetime is assumed to be a linear function of the cell damage at the division $y$. Because $y$ is determined by the damage at birth $x$, one can generally write $\tau(x)=\tau_0+\mu y(x)$.
\item Chao model \cite{chao2010model}. Here it is assumed that the lifetime is the time when some product $P$, 
whose synthesis rate depends on the damage, $\dot P=1-s D(t;x)$ with $P(t_0)=0$, reaches a threshold value $P_0$. This leads to a nontrivial dependence of the cell lifetime on its initial damage $\tau(x)$, see Eq.~\eqref{eq:chtim} below.
\end{enumerate}
\item  \textbf{Damage inheritance}:  A deterministic rule according to which the damage of a 
mother cell is distributed between two daughters (no damage losses),
$x_1=f(y),\ \ x_2=y-f(y),\quad 0\leq f(y)\leq y$.

{\em Examples:}
\begin{enumerate}[a.]
\item In model~\cite{chao2010model}, a linear mapping $x_{1,2}=\frac{1\pm a}{2}y$ with a constant asymmetry parameter $0<a<1$ is adopted.
\item In model~\cite{vedel2016asymmetric}, a nonlinear mapping  $x_{1,2}=\frac{1\pm a(y)}{2}y$ with the asymmetry
parameter being a function of the damage of the mother cell, $a(y)$, is suggested.
\end{enumerate}
\end{enumerate}
 All model variants listed above are deterministic, however we will generalize our description to account for stochasticity in ADS in Sec. \ref{sec:stoc}.

\section{Qualitative description of the deterministic cell population dynamics}
\label{sec:chaos}

\begin{figure}
\centering
\includegraphics[width=0.49\textwidth]{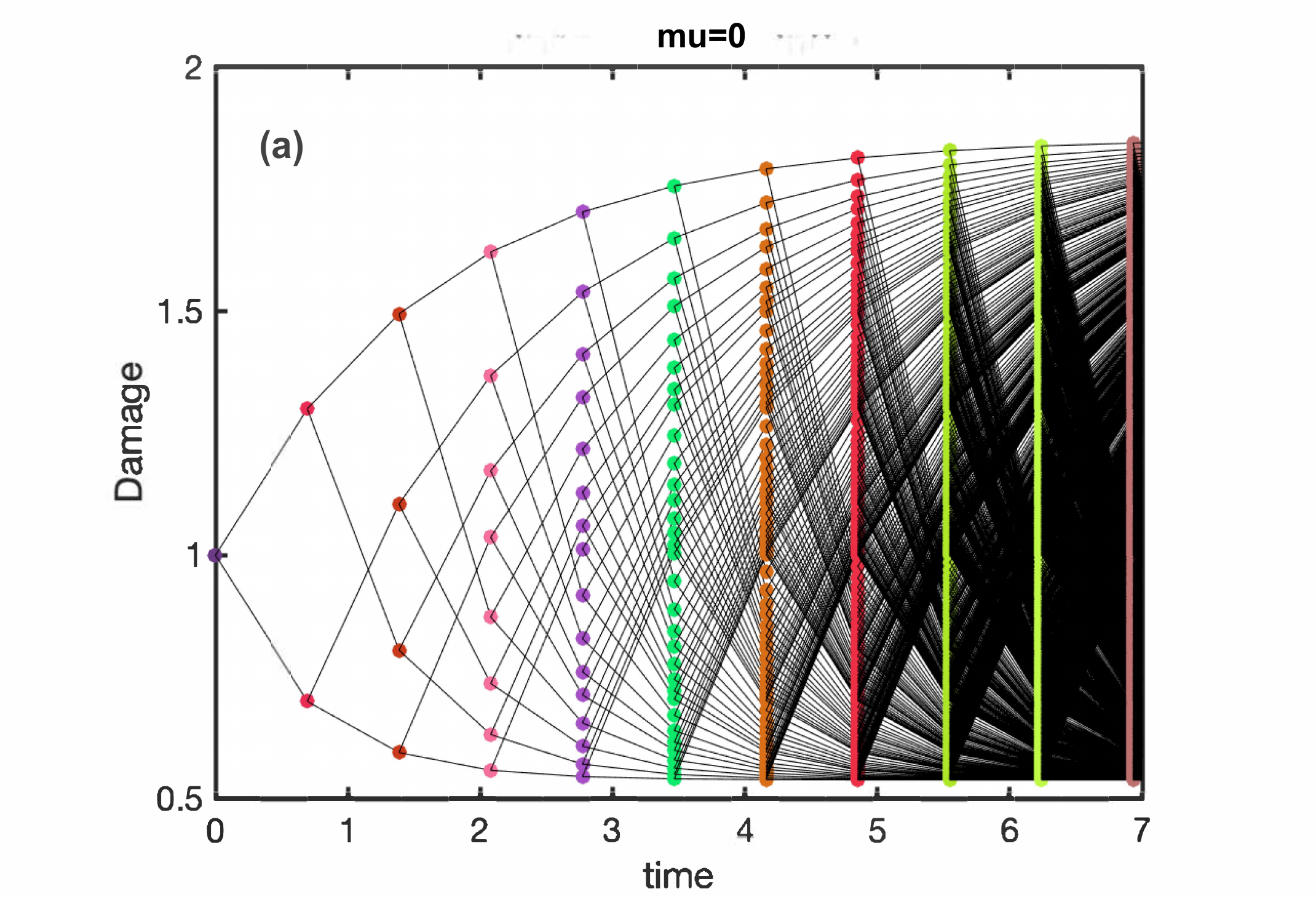}\hfill
 \includegraphics[width=0.49\textwidth]{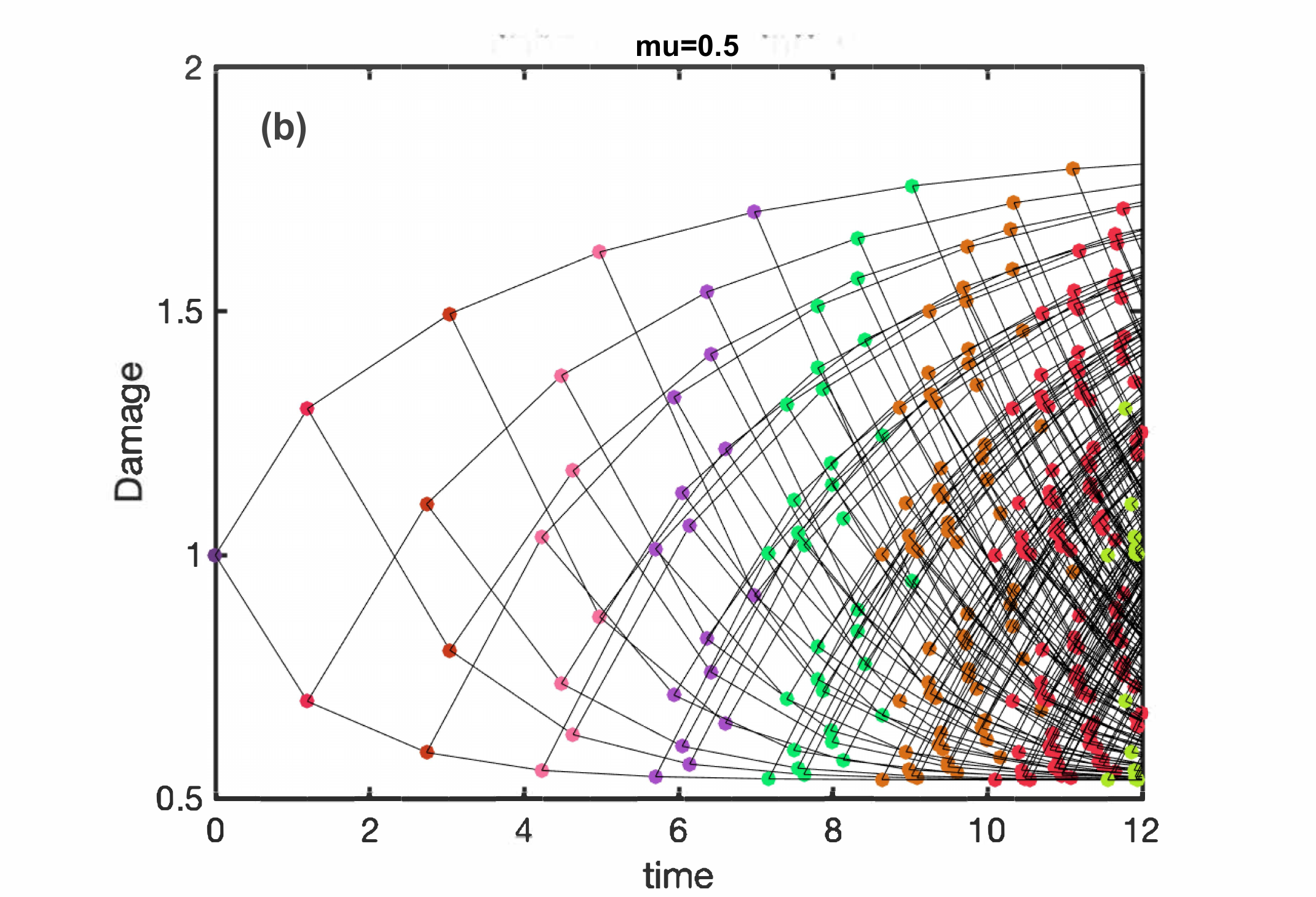}\\
\includegraphics[width=0.49\textwidth]{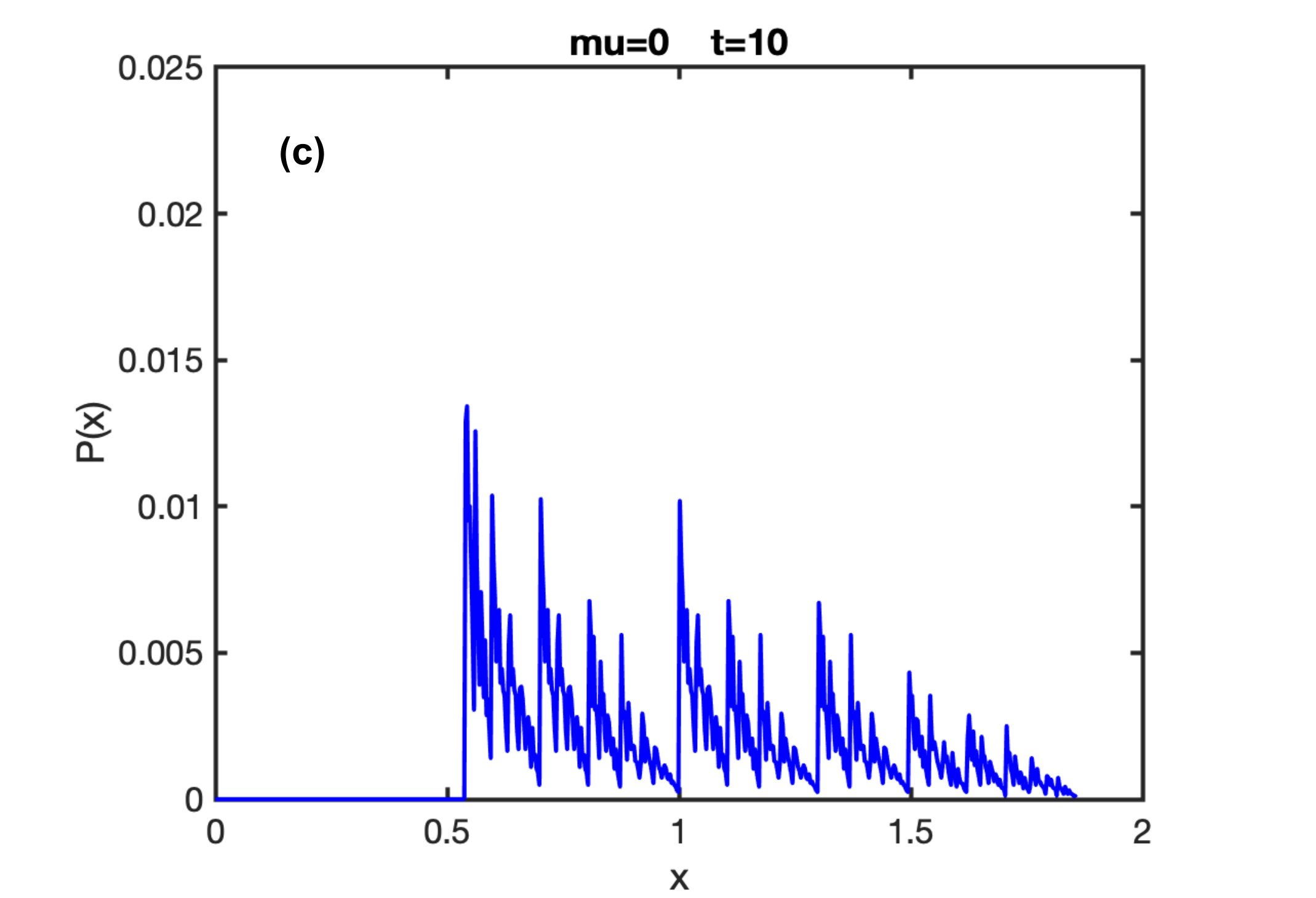} \hfill
\includegraphics[width=0.49\textwidth]{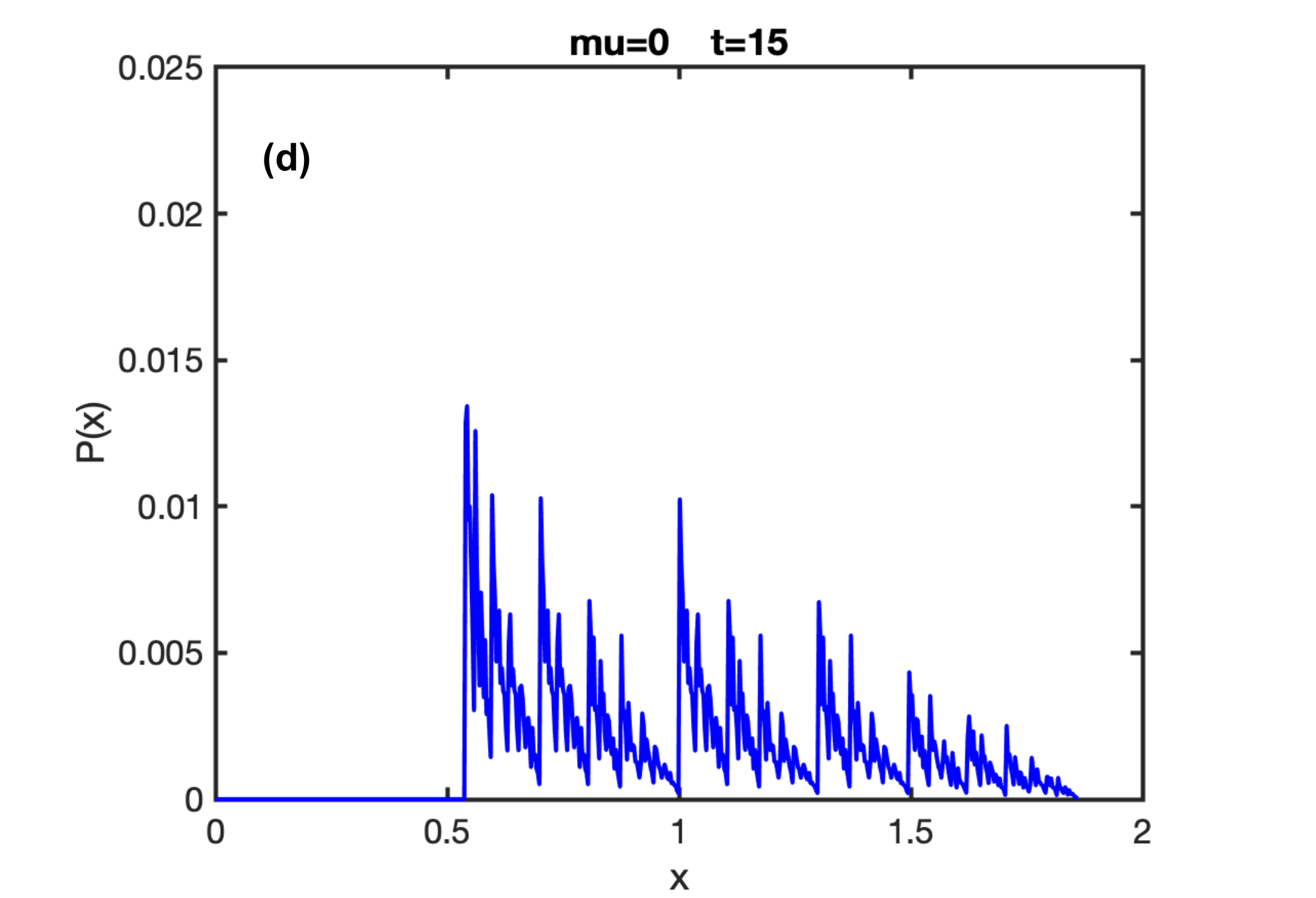}\\
\includegraphics[width=0.49\textwidth]{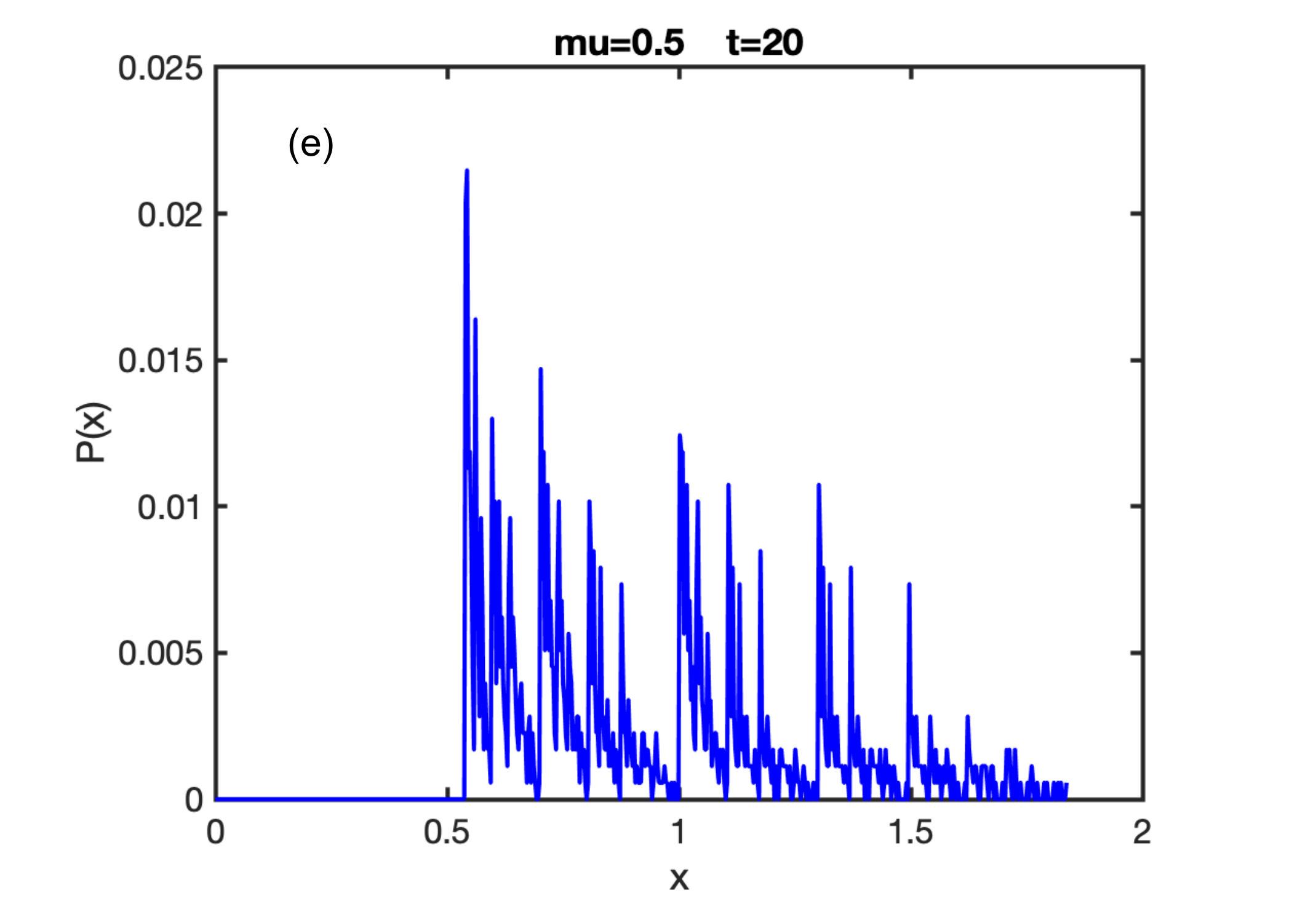} \hfill
\includegraphics[width=0.49\textwidth]{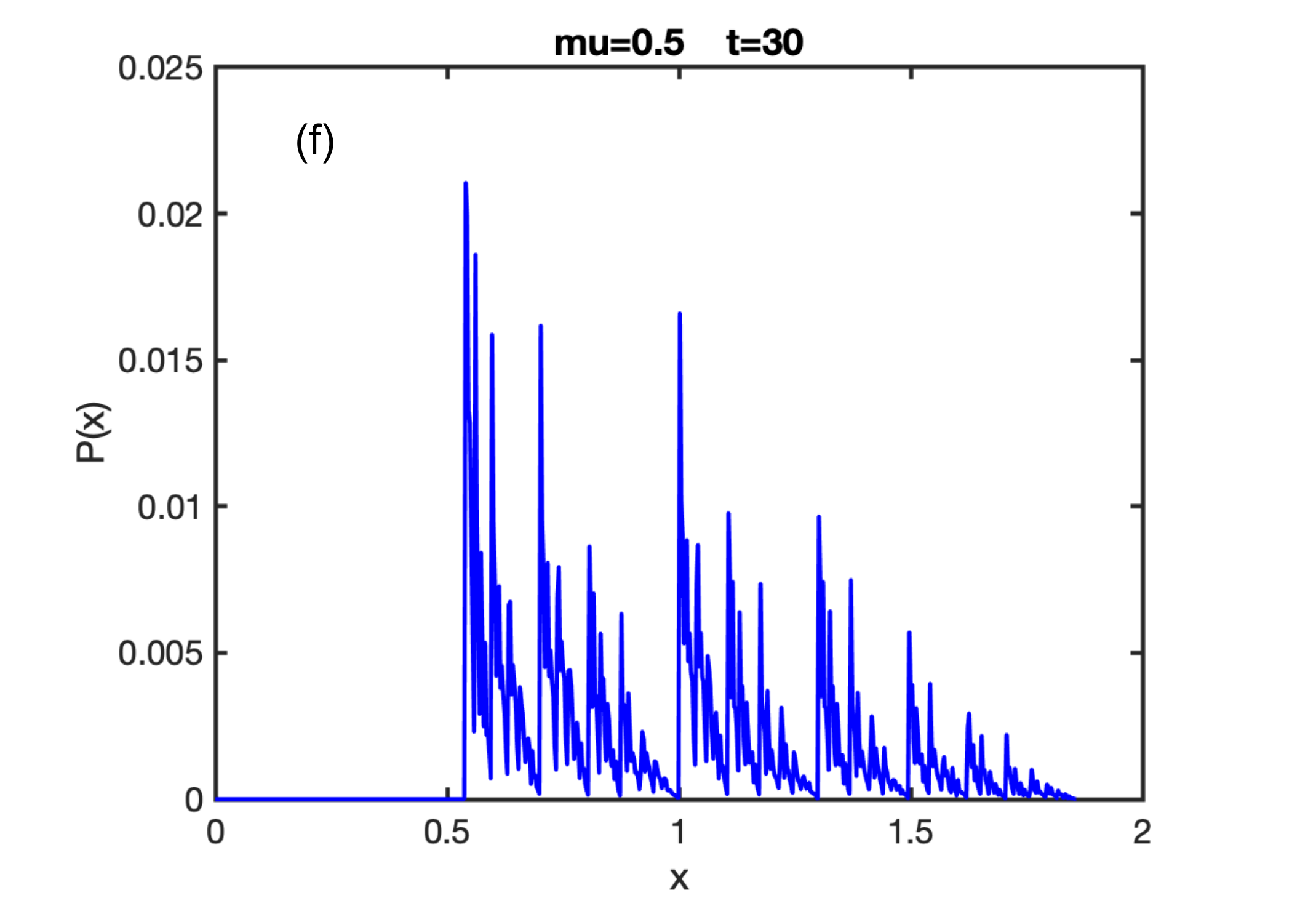}
\caption{Simulations of damage propagation according to model 1a2a3a starting from a single cell. In top two panels each dot represents initial damage of a cell, and ridges represent mother-daughter ancestry. Dots of the same color correspond to cells of the same generation. The parameters are  $a=0.3, \lambda=1, \tau_0=\ln 2$, and $\mu=0$ (a) or $\mu=0.5$ (b). Panels (c) and (d) show the distributions of damage in just-born cells for for the simulation (a) at times $t=10$ (c) and $t=15$ (d). Similarly, panels (e) and (f) show the distributions of damage in just-born cells for the simulation (b) at times $t=20$ (c) and $t=30$ (d).  Comparing panels (c) with (d) and (e) with (f) we observe that the process quickly to the asymptotic long-time regime. A detailed characterization of the early transients is beyond the scope of this paper.
}
\label{fig:scatter}
\end{figure}

It is straightforward to simulate any specific ADS model numerically, starting, for example, from a single cell with a certain initial 
damage. Fig.~\ref{fig:scatter}a,b illustrate these dynamics  for a model with the fixed damage gain (rule 1a of the previous Section), 
linear lifetime function (2a), and linear damage inheritance rule (3a). In the following we will label this model as 1a2a3a. 
Fig.~\ref{fig:scatter}a corresponds to the constant lifetime ($\mu=0$) and  Fig.~\ref{fig:scatter}b to 
the damage-dependent lifetime ($\mu > 0$). Note that since the number of cells in such a simulation grows exponentially with time, 
it cannot be extended very far. As an alternative, one can simulate a Moran-type process with a constant population of $N$ cells when 
at each cell division, another random cell is removed from the population, so the population size remains constant. The statistical 
features of both processes are the same in the ``thermodynamic limit'' $N\to\infty$, where one can neglect finite-size fluctuations. We can also expect them to be similar for finite $N$  at the time when the size of a growing population is also $N$.

The main difference between the cases in panels (a) and (b) is that for the damage-independent lifetime  
all cells divide simultaneously, while for damage-dependent lifetimes, synchrony of 
division times is quickly lost.   In both cases, after a short transient, the distribution of 
damage in just-born cells appears to reach a broad and multi-peaked stationary state (Fig.~\ref{fig:scatter}c,d) with a finite support between certain $x_{min}$ and $x_{max}$. It is straightforward to find $x_{min}$ and $x_{max}$ for a given model as the asymptotic limit of inheriting only smaller or only larger fractions of mother's cell damage within a lineage. For example, for Model 1a2a3a, $x_{min}=\lambda\frac{1-a}{1+a}$, $x_{max}=\lambda\frac{1+a}{1-a}$.

It is, however, not trivial  to interpret these simulations. 
Indeed, the process is fully deterministic: an initial damage of a cell fully  determines 
damages and fission times of all its descendants. But it is not a classical deterministic 
dynamical system because the number of cells grows, and we cannot represent the time 
evolution of the damage values in all descendant cells as a single trajectory. However, 
while each mother cell has two daughters,  {the daughter cell damage uniquely specifies the 
damage of its mother, at least for the model 1a2a3a, if parameters $a$ and $\lambda$ are also known. The reason for this is that the damages of daughters receiving larger and smaller fractions of the mother's damage belong to non-overlapping sets $(x_{min},\lambda)$ and $(\lambda,x_{max})$ (this is not the case for all ADS models, see below). 
Thus, for the model 1a2a3a we can construct a deterministic `back-in-time' map for the damage $x(n)$ of the just-born cells in $n$-th generation 
\begin{equation}
x(n-1)=
\begin{cases}
\frac{2}{1-a} x(n)-\lambda &\text{for }x_{min}\leq x(n)\leq\lambda\;,\\
\frac{2}{1+a} x(n)-\lambda &\text{for }\lambda<x(n)\leq x_{max}\;,
\end{cases} 
\label{eq:bermap}
\end{equation}
(Fig.~\ref{fig:traj}a) and generate a unique ``lineage''  damage trajectory back in time, as illustrated by Fig.~\ref{fig:traj}b 
where we concatenated the damage time courses in different generations. 

\begin{figure}
\centering
\includegraphics[width=\textwidth]{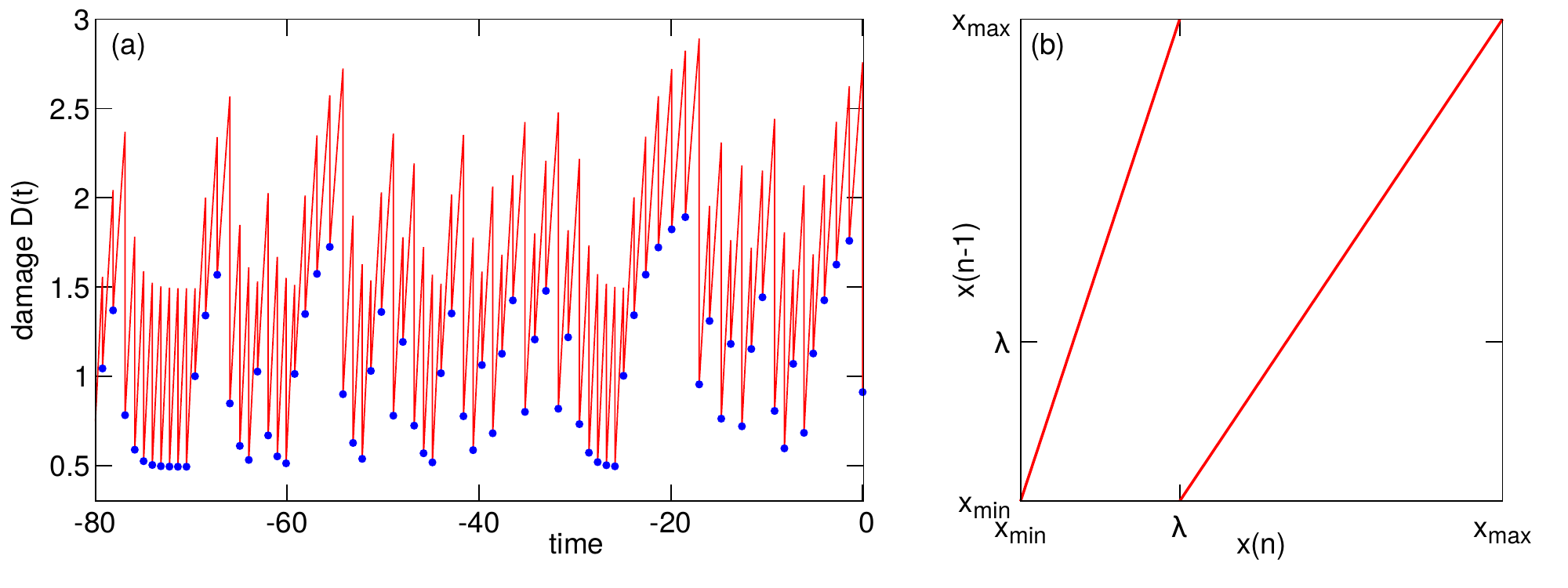}
\caption{Panel (a): a concatenated trajectory of damage in the lineage of ancestors of 
a given cell in Model 1a2a3a.  Parameters: $A=0.33$, $B=0.67$,  $\lambda=1$, $\tau_0=\ln(2)$, $\mu=0.4$. 
Red line: damage vs time $D(t)$;
blue dots: damages of just-born cells $x$ in subsequent generations.
Panel (b): a back-in-time map of initial damages in mother vs daughter cell for model 1a2a3a.
}
\label{fig:traj}
\end{figure}

The main conclusion of Fig.~\ref{fig:traj}b is that the deterministic nature of the model notwithstanding, the time course of the damage is highly irregular, which suggests
that this process may be approached from the viewpoint of the chaos theory.  The mapping \eqref{eq:bermap} is indeed an expanding piecewise linear map which is a prototypical model of deterministic chaos~\cite{ott2002chaos}. }This immediately explains the irregularity of the lineage damage trajectory. 
In fact, relation \eqref{eq:bermap} can be considered a Poincar\'e map for the 
continuous trajectories of the full time-dependent damage dynamics. The latter is described not by differential
equations, but rather by a combination of a continuous evolution of damage during the cell lifetime
interrupted by a discrete transition from mother to daughter by the rule of damage inheritance. In this sense
our hybrid continuous-discrete system resembles an integrate-and-fire model for firing neurons,
where a continuous voltage increase (current integration) is combined with a jump rule (firing of a spike
and a reset of the voltage).

We stress here that although a back-in-time trajectory can be always extracted from a
forward-in-time simulation of a population of dividing cells, not in every case this back-in-time trajectory can be associated with a deterministic dynamical system
like the map \eqref{eq:bermap}. It requires the uniqueness of the back-in-time map, i.e. for any initial damage of a daughter cell there should be only one possible value
of the mother's  initial damage. A detailed analysis (Sec. \ref{sec:flt}) shows that this is not always the case.  For example, for model 1b2a3a with $\beta>0$, two branches of the back-in-time map similar to Fig.~\ref{fig:traj}b) overlap, and so neither forward-in-time, nor back-in-time lineage trajectories can be computed by iterating a deterministic 1D map. 

\begin{figure}
\centering
\includegraphics[width=1\textwidth]{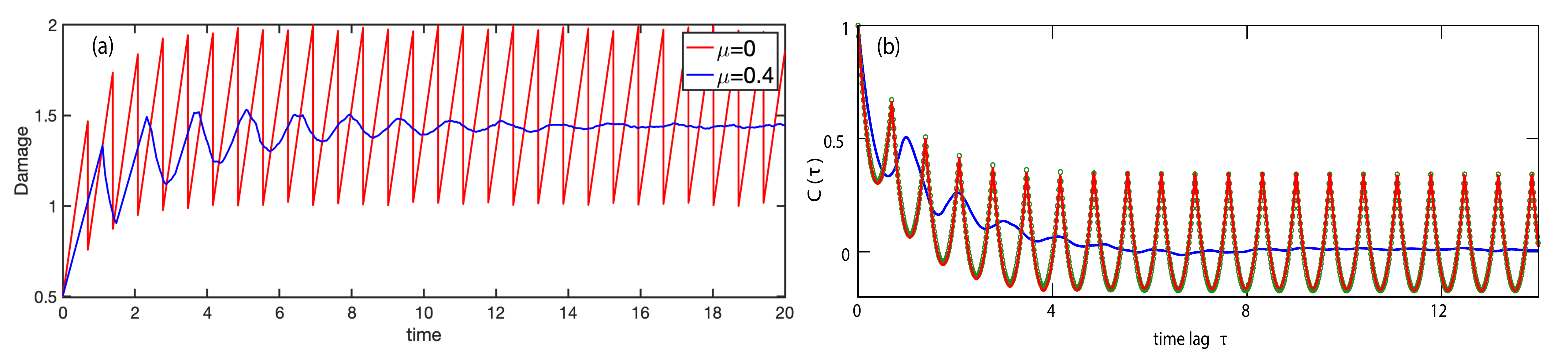}
\caption{(a) Mean damage of a population of 5000 cells undergoing a Moran process for  ADS model 1a2a3a 
for $\mu=0.4$ (blue) (blue) and for $\mu=0$ (red).  Other parameters are the same as in Fig.~\ref{fig:traj}. 
(b) The autocorrelation function for a single back-in-time damage trajectory and the same parameters as in (a).
  Green dots: theoretical  expression for the autocorrelation function for constant lifetimes (see \ref{ap:cf}). }
\label{fig:acf}
\end{figure}

The essential ingredient of any ADS model is the dependence of cell lifetime on its damage. Continuing with our analogy between the mother-to-daughter damage inheritance and a Poincar\'e map, one can interpret the lifetimes as 
Poincar\'e return times. The properties of these times affect the regularity of the damage
trajectories Fig.~\ref{fig:traj}a; the corresponding notion in the chaos theory 
is {\em phase coherence}~\cite{Pikovsky-Rosenblum-Kurths-01}. One may distinguish a generic case of damage-dependent 
lifetimes from a degenerate case of fixed lifetimes. For fixed lifetimes, the time evolution of the damage is only partially irregular: 
Divisions occur at regular time intervals (Fig.~\ref{fig:scatter}a), and there is no mixing for the continuous-time dynamical process. 
All lineages starting from the same cell undergo divisions simultaneously, and so the mean population damage continues to oscillate 
indefinitely (see Fig.~\ref{fig:acf}a, red line).  In other words, all the cells remain ``in phase''.
In contrast, for a generic damage dependence of lifetimes, the intervals between sequential divisions are chaotic (since the initial 
damages of cells in a lineage are chaotic), and the continuous-time process is mixing (Fig.~\ref{fig:scatter}b). Thus, different 
lineages starting from the came cell decorrelate (in other words, the phases of different cells become scattered)
and the mean population damage eventually settles into a steady state (see Fig. \ref{fig:acf}a, blue line). Another way to 
quantify this difference is to compute the normalized autocorrelation function of the damage trajectory
\begin{equation}
C(\tau)=\frac{\av{(D(t)-\av{D})(D(t+\tau)-\av{D})}}{\text{Var}(D)}\;,
\label{eq:corf}
\end{equation}
 {where the angular brackets denote averaging over different lineages
\footnote{ {For averaging we used here so-called retrospective (or historic) sampling  by taking lineages of all cells present at a given time (see \cite{nozoe2017inferring,thomas2017making}). For an alternative  chronological (or forward) sampling), when each daughter of a given cell is selected with equal probability, the averages may be slightly different since it overestimates older, slower-dividing cells}}.}

In Fig.~\ref{fig:acf}b we compare this autocorrelation function for model 1a2a3a with $\mu\ne 0$ and $\mu=0$. One can see that while for damage-dependent lifetimes ($\mu\ne 0$)  this function decays to zero, for constant lifetimes it initially decays (reflecting the irregularity of the damages in just-born cells), but asymptotically at large
time lags it oscillates periodically without decay, reflecting phase coherence of the corresponding damage trajectory. For the latter case,  the autocorrelation function can be computed analytically (see \ref{ap:cf}). 

\section{Kinetic description of the cell population dynamics} 
\label{sec:kin}

In the previous section we demonstrated that the dynamics of the asymmetric damage segregation in cell lineages is irregular, thus it appears appropriate to describe them statistically.   In this section, we formulate a kinetic description of ADS in the thermodynamic limit of a large number of cells. It is similar to the recently developed transport approach \cite{levien2021non,min2021transport} but differs in some details. 

As we have argued above, in many aspects the damage dynamics is analogous to the dynamics of chaotic oscillators, and we will follow this analogy in the construction of the statistical theory. It is convenient to introduce a phase of the cell cycle $\phi$ that changes linearly from $\phi=0$ (birth) to $\phi=1$ (division) with the rate $\w$  (which is simply the inverse of the cell lifetime $\tau$) that generally may depend on the initial cell damage $x$:
\begin{equation}
\frac{d\phi}{dt}=\w(x)=\frac{1}{\tau(x)}\;.
\label{eq:phd}
\end{equation}

Next, instead of $D(t;x)$ above, we can introduce the damage during the cell lifetime as a deterministic function of the initial state $x$ and the phase $\phi$,  
$D=F(x,\phi)$. This function is related to the function $D(t;x)$ above as
\begin{equation}
F(x,\phi)=D(\tau(x)\phi;x),\qquad F(x,0)=x,\qquad y=F(x,1).
\label{eq:cd} 
\end{equation}

Let us now introduce a two-dimensional population density $N(x,\phi, t)$, so that
$N(x,\phi,t)dx d\phi$  is the number of cells in a small interval of the initial damages $dx$
 and in a small interval of phases $d\phi$. With a proper normalization, the quantity $N(x,\phi,t)$
 can be interpreted as a probability density in the space of deterministic variables $x,\phi$. We stress
 here that the irregularity of the dynamics is not needed for this:  even regular deterministic processes (e.g. populations of periodic oscillators) can be described via a probability density in the phase space, with an additional assumption of random initial phases in an ensemble of copies \cite{kuramoto1984chemical}. This quantity is similar to the distribution $\psi(u,x,t)$ introduced in \cite{min2021transport}, which instead of our phase $\phi$ depends on the actual cell age $u$ measured in units of time. We believe that our distribution of cell ages in terms of the phase $\phi$ (i.e. the fraction of the cell's lifetime) has its advantages, especially when the lifetime varies strongly from cell to cell. 
 
 Instead of dealing with an exponentially growing population of dividing cells, it is convenient to 
 introduce cell death, so that a certain fraction of cells $\Gamma N(x,\phi, t)dt$ dies within a small time interval $dt$
 irrespective of $x$ or $\phi$.  The death rate $\Gamma$ in principle can be arbitrary, but if we choose it exactly 
balance the (yet unknown) average asymptotic proliferation rate, the total number of cells will remain constant, and the distribution $N(x,\phi,t)$ should eventually 
 approach a stationary state $N(x,\phi)$. It is easy to see that the death rate $\Gamma$ found from the stationarity condition of the asymptotic solution for a model with death is exactly the growth rate of a proliferating cell population without death. Alternatively, we can assume a Moran-type 
 process~\cite{moran1958random} when at the time of every binary fission, one other randomly chosen cell is taken away, so the total number of 
 cells remains precisely constant at all times, not just asymptotically and in the thermodynamic limit.  
 It can be shown (see \ref{ap:moran}) that asymptotically in the large-time limit the kinetic equation for the Moran-type process is equivalent to the one presented here. 

The transport  equation for $N(x,\phi,t)$ in the system with cell death follows from the probability conservation: 
\begin{equation}
\frac{\partial N}{\partial t}+\frac{\partial}{\partial \phi}[N\w(x)]=-\Gamma N,
\label{eq:ke}
\end{equation}
where the second term on the l.h.s. describes cells advection from phase 0 to phase 1 with  phase velocity $\w(x)$ according to \eqref{eq:phd},
 and the r.h.s. term describes cell death.  {This equation formally coincides with the transport equation for the normalized damage distribution in an exponentially growing population \cite{min2021transport} with $\Gamma$ playing the role of the population growth rate instead of the death rate in our model.} We assume that the death rate $\Gamma$ is indeed chosen properly 
 to balance the population growth. In this case, and if the initial phases in a population are random, the probability distribution asymptotically approaches a stationary state described by the equation~\footnote{If the dynamics are mixing, the initial random phase approximation is not required, and the stationary asymptotic distribution will be achieved from an arbitrary initial distribution, but for the degenerate case of damage-independent lifetimes, initial ``randomization'' is necessary.}
 \begin{equation}
\frac{\partial}{\partial \phi}[N\w(x)]=-\Gamma N\;.
\label{eq:station}
\end{equation}
The solution of this equation describes the exponentially decaying flux of cells from $\phi=0$ towards $\phi=1$:
\begin{equation}
N(x,\phi)=N(x,0) \exp[-\Gamma\tau(x)\phi].
\label{eq:nphi}
\end{equation}
To close the system and to ensure stationarity, we need to add  the boundary (transition) condition balancing the incoming flux of the daughter cells $N(\bar{x},0) \w(\bar{x})$ at $\phi=0$ with the outgoing flux of mother cells $N(x,1)\w(x)$ at $\phi=1$ (due to the cell division, the number of cells leaving the interval at $\phi=1$ has to be exactly the half of the number of cells entering at $\phi=0$, thus factor 2 in Eq.~\eqref{eq:Gamma1} below) . 

The connection is determined by the rule specifying how the damage at the end of the life cycle $y=F(x,1)$ is redistributed
within daughter cells.  For a general damage inheritance rule (see 3. in section \ref{sec:models}) we have
\begin{equation}
\begin{aligned}
\bar{x}_1&=g_1(x)=f(F(x,1))\equiv f_1(F(x,1))\\
\bar{x}_2&=g_2(x)=F(x,1)-f(F(x,1))\equiv f_2(F(x,1))
\end{aligned}
\label{eq:genmap}
\end{equation}
For monotonous functions $F$ and $f$, depending on  their specific forms,  a daughter cell with initial damage $\bar{x}$ can descend from one of two mothers with initial damages $x_{1,2}=F^{-1}(f_{1,2}^{-1}(\bar{x}),1)$. 

The balance condition for the fluxes thus can be written as
\begin{equation}
\w(\bar x) N(\bar x,0)\;d \bar x=\sum_{i=1}^2
\w(x_i)N(x_i,1) dx_i\equiv  \sum_{i=1}^2 \w(x_i)N(x_i,0)e^{-\Gamma \tau(x_i)} dx_i.
\label{eq:bc}
\end{equation}
Dividing both sides by $dx$ we arrive at the basic Frobenius-Perron operator for the damage distribution in a population of dividing cells at the beginning of their cell cycle 
\begin{equation}
 {
\w(\bar x) N(\bar x,0)=\sum_{i=1}^2\w(x_i)N(x_i,0)e^{-\Gamma \tau(x_i)}\left[{d\over dx_i} f_i(F(x_i,1))\right]^{-1}.
}
\label{eq:bc1}
\end{equation}
Alternatively, we can write this Frobenius-Perron equation for the product $P(x)=\w(x)N(x,0)$ in the following integral form
\begin{equation}
P(x)=\int dx' [\delta(x-g_1(x'))+\delta(x-g_2(x'))]P(x')e^{-\Gamma \tau(x')}
\label{eq:bc2}
\end{equation}
where $g_{1,2}(x)=f_{1,2}(F(x,1))$ and we dropped the bar over $x$. It is easy to see that properly normalized $P(x)$ is the density of just-born cells (the fraction of the whole population of cells that that are born within a small unit time interval with initial damage $x$). 

This equation can also be deduced directly from the general equation for the stationary initial damage distribution function derived in \cite{min2021transport}
\begin{equation}
P(x)=2\int dx'g(x|x')P(x')e^{-\Gamma \tau(x')}\;,
\label{eq:bcg}
\end{equation}
where $f(x|x')$ is the probability to have initial damage $x$ in a cell if the initial damage in its mother cell is  $x'$. Eq.~\eqref{eq:bcg} can describe both deterministic and stochastic damage distribution scenarios. To obtain our deterministic \eqref{eq:bc2} one just needs to take $g(x|x')={1\over 2}[\delta(x-g_1(x'))+\delta(x-g_2(x'))]$.

The conservation of the number of cells requires 
\begin{equation}
\begin{gathered}
\int dx\;\w(x)N(x,0)=\sum_{i=1}^2 \int\;dx_i \; \w(x_i)N(x_i,0)e^{-\Gamma \tau(x_i)}\equiv\\
 2 \int\;dx \; 
\w(x)N(x,0)e^{-\Gamma \tau(x)}\;,
\end{gathered}
\label{eq:Gamma0}
\end{equation}
or
\begin{equation}
\int dxP(x)=2 \int\;dx P(x)e^{-\Gamma \tau(x)}\;.
\label{eq:Gamma1}
\end{equation}
This self-consistency condition  together with Eq.~(\ref{eq:bc2}) uniquely determines the death/growth rate $\Gamma$. 

Asymptotically, the solution of the Frobenius-Perron equation \eqref{eq:bc1} (provided
the death rate $\Gamma$ satisfies \eqref{eq:Gamma0}) 
converges to the stationary damage distribution at the moments of birth $N(x,0)$. 
It is easy to see that this distribution has a finite support between $x_{min}$ and $x_{max}$ which are determined from equations
$x_{min}=g_1(x_{min}),\ x_{max}=g_2(x_{max})$ (we assume for definiteness that $g_1(x)<g_2(x)$).

Let us introduce the total number of cells (here we use also \eqref{eq:nphi}):
\begin{equation}
\mathcal{N}=\int dx\;d\phi\; N(x,\phi)=\int dx\;d\phi\; N(x,0)\exp[-\Gamma\tau(x)\phi]\;.
\label{eq:nc}
\end{equation}
Then the stationary probability distribution density over initial damages and phases is
\begin{equation}
n(x,\phi)=\frac{1}{\mathcal{N}}N(x,\phi)=\frac{1}{\mathcal{N}}N(x,0)\exp[-\Gamma\tau(x)\phi]\;.
\label{eq:dd}
\end{equation}
We can use this two-dimensional density to obtain the  density for the distribution
of the damages (now not initial one, but just observed at all possible phases), by the following expression 
\begin{equation}
\begin{gathered}
W(D)=\av{D-F(x,\phi)}=\int dx\;\int_0^1 d\phi\; \delta(D-F(x,\phi)) n(x,\phi)=\\=
\frac{1}{\mathcal{N}}
\int dx\; \int_0^1 d\phi\; \delta(D-F(x,\phi)) N(x,0)\exp[-\Gamma \tau(x)\phi]\;.
\end{gathered}
\label{eq:wd}
\end{equation}
Here we used expression \eqref{eq:cd} which relates the damage at the phase $\phi$ to the initial damage $x$. 
We can also obtain he distribution over the phases $w(\phi)$  as the marginal distribution by integrating
the two-dimensional density \eqref{eq:dd}
\[               
w(\phi)=\int dx\;\frac{1}{\mathcal{N}}N(x,\phi)=\mathcal{N}^{-1}\int dx
N(x,0)\exp[-\Gamma\tau(x)\phi]\;.
\] 


\section{Linear damage accumulation, constant lifetime}
\label{sec:flt}


In this section we use the kinetic theory developed in the previous section to analyze the statistical properties of the model with a fixed lifetime $\tau=T$ and linear mappings
\begin{equation}
g_1(x)=A(x+\lambda),\qquad g_2(x)=B(x+\lambda)
\label{eq:linmap}
\end{equation}
with constant $A$ and $B$ where for definiteness, we assume $A<B$. 

It is easy to see that linear transformations (\ref{eq:linmap}) with $A+B=1$ correspond to rules 1a and 3a of Sec. \ref{sec:models}, where 
\[
A=\frac{1-a}{2},\quad B=\frac{1+a}{2}. 
\]
The same linear mappings also can be deduced from rule 1b, however in this case, integration of the ODE for the damage within the cell cycle yields 
\[
A=\frac{1-a}{2}e^{\beta T},\quad B=\frac{1+a}{2}e^{\beta T},\quad \lambda= \gamma\beta^{-1}(\exp[\beta T]-1),
\]
and so for non-zero $\beta$, $A+B\ne 1$. If $\beta<0$, then $A+B<1$, and if $\beta>0$, then $A+B>1$. Note that for the damage to remain bounded at all times it is required that $A<1, B<1$, i.e. $A+B<2$.

The fixed lifetime may correspond either to rule 2a with $\mu=0$ (then $T=\tau_0$) or to rule 2b with $s=0$ (then $T=P_0$).

\subsection{Fractal properties of the damage distribution}
\label{sec:ifs}

 The discrete dynamics of $x$ according to two linear transformations \eqref{eq:linmap} is a well-known mathematical object called \textit{Iterated Function System} (IFS), see 
 \cite{peres2006absolute,shmerkin2019furstenberg,barral2021multifractal}.  The particular case $A=B$
 is often called Bernoulli convolution~\cite{peres2000sixty}. IFS 
 is the simplest mathematical model generating fractals~\cite{barnsley2014fractals}. To illustrate this, we first write down the equation for $P(x)$ that follows from the general Frobenius-Perron equation \eqref{eq:bc2}:
 \begin{equation}
 P(x)=\frac{1}{2A}P\left(\frac{x}{A}-\lambda\right)+\frac{1}{2B}P\left(\frac{x}{B}-\lambda\right)\;,
\label{eq:fp0}
\end{equation}
where we also took into account the condition $e^{\Gamma T}=2$ which immediately follows from \eqref{eq:Gamma1} for $\tau(x)=T$. 
The solution of (\ref{eq:fp0}) is localized between $x_{max}=B\lambda/(1-B)$ and $x_{min}=A\lambda/(1-A)$. 

\begin{figure}
\centering
\includegraphics[width=\textwidth]{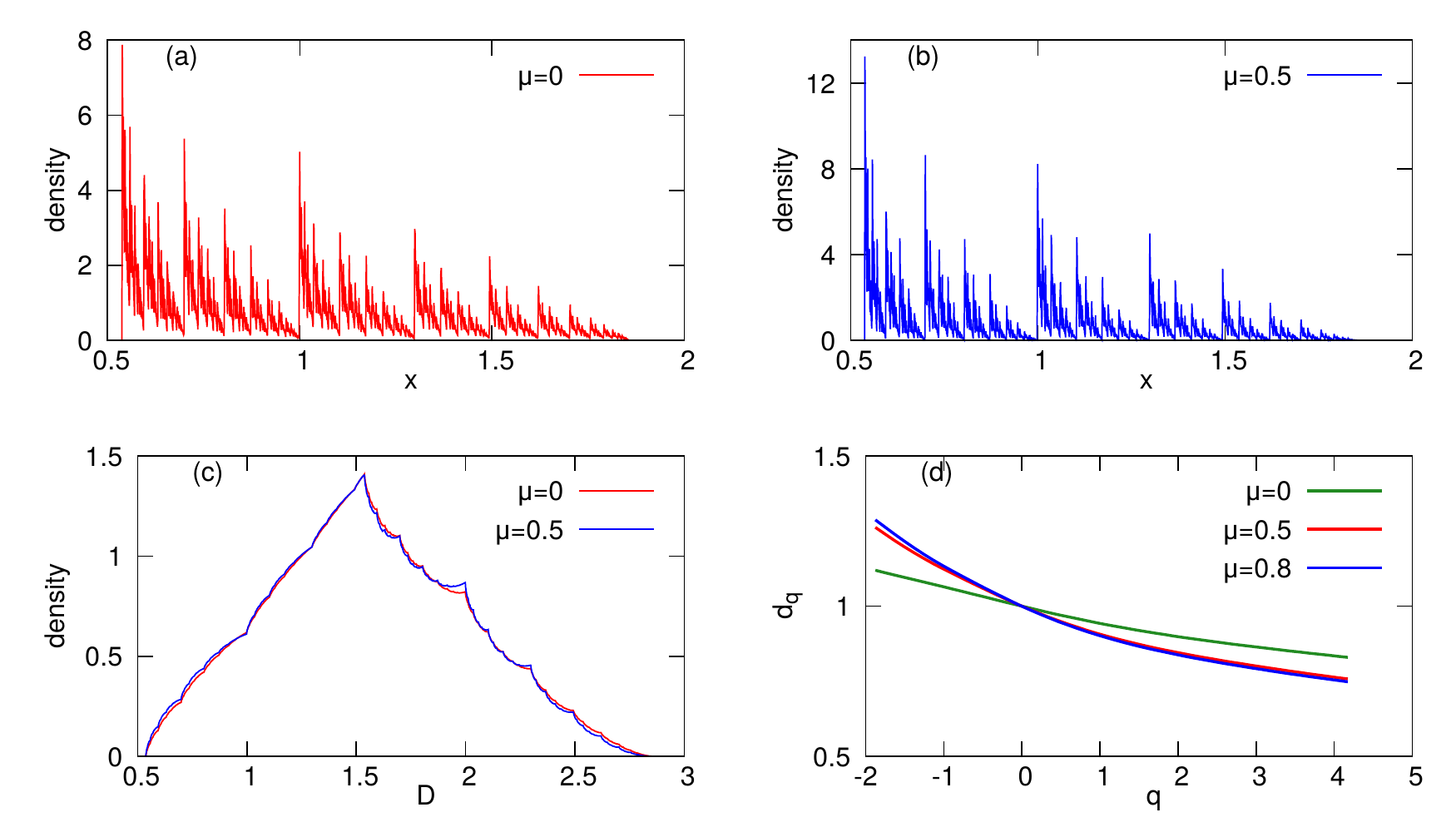}
\caption{Panel (a): Solution $P(x)$ 
of eq.~\eqref{eq:fp0} for $A=0.35$, $B=0.65$ and $\lambda=1$, $\tau_0=\ln(2)$
obtained via discretization of the FP operator. Panel (b):  Solution $P(x)$ in the case
of damage-dependent lifetimes (Eq.~\eqref{eq:frobp1}), with $\mu=0.5$. These two fractal distributions
correspond to those in Fig.~\ref{fig:scatter}. Panel (c): Distributions of damage $D$ for the total population, according to
expression \eqref{eq:wd}, for the densities in panels (a,b). Panel (d): Spectra of generalized fractal dimensions for different $\mu$.
}
\label{fig:f1}
\end{figure}

Numerical iteration of the operator \eqref{eq:fp0} for arbitrary $A$ and $B$  generically yields a fractal distribution, as exemplified by Fig.~\ref{fig:f1}a where we used the same parameters as in Fig.\ref{fig:scatter}a \footnote{The peaks of this distribution are infinite, they appear finite in Fig.~\ref{fig:f1} because in numerical iterations we used a finite number of bins.}. This distribution is virtually indistinguishable from the distribution obtained in direct simulations of the underlying ADS model (Fig.\ref{fig:scatter}c). 

The fractal properties of this distribution can be summarized as follows (see~\cite{barral2021multifractal} for the current state of the theory).

\paragraph{\textbf{Case without overlap $A+B\le 1$.}} This is the simplest case where there is no overlap 
of two linear branches in the mapping \eqref{eq:linmap}. In this case one can characterize the invariant 
measure by generalized fractal dimensions $d_q$ (see, e.g., \cite{ott2002chaos}; traditionally for these 
dimensions capital letters are used, but in this paper this notation is reserved for the damage). 
A standard scaling argument~\cite{ott2002chaos} leads to the exact parametric expression 
\begin{equation}
d_q=\frac{\mathcal{T}}{q(\mathcal{T})-1},\quad q(\mathcal{T})=\frac{\ln (A^{-\mathcal{T}}+B^{-\mathcal{T}})}{\ln 2}\;.
\label{eq:frdim}
\end{equation}
Most important are the box-counting dimension $d_0$ and the information dimension $d_1$. One can see that
when the damage of just-born cells is conserved $A+B=1$, the box-counting dimension is one, i.e. the support of the measure is the full interval $[x_{min},x_{max}]$. In other words, there are no voids and the set of possible damages is not a classical Cantor set. In contrast, if the initial damage is partially dissolved, i.e. $A+B<1$, then 
$d_0<1$ and the set of possible damages is a Cantor set.  In both cases the information dimension $d_1<1$ (with the exception of a trivial degenerate situation of symmetric segregation $A=B=1/2$, when the measure is uniform).

\paragraph{\textbf{Case with overlap $A+B>1$.}} In this case the two branches of \eqref{eq:linmap} overlap. This situation has long been a conundrum for the measure theory, and only recently it has been partially clarified  \cite{peres2006absolute,shmerkin2019furstenberg,barral2021multifractal}. In particular, Ref.~\cite{barral2021multifractal} analyzed generalized dimensions in the range $0\leq q\leq 1$. 
For typical $AB>1/4$, all these dimensions are $d_q=1$, what means that the distribution is continuous with a finite density $P(x)$. For $AB<1/4$ and $A+B>1$, there is a ``phase transition'' in dependence on $q$: there is a critical value $q^*$, beyond which the expression \eqref{eq:frdim} holds, while below this value $D_q$ is a fractional-linear function of $q$, with $d_0=1$.

A more detailed understanding exists for the symmetric case $A=B$, which is called Bernoulli convolution.
Here, it has been proven  that for almost all values of $A>1/2$, the invariant measure is absolutely continuous 
\cite{solomyak1995random}. At exceptional points (so-called  Pisot numbers) the distribution is fractal, but the information dimension
is very close to one (see recent estimates in \cite{feng2021estimates,kleptsyn2022uniform}). We refer to 
Ref.~\cite{bandt2018finite} for a nice illustration of the densities for different values of $A>1/2$.

\subsection{Singularities at the ends of the interval}
One can see from Fig.~\ref{fig:f1} that the behavior of the damage distribution density is very different at the opposite ends of the interval, i.e. close to $x_{min}$ and  $x_{max}$. While there is a sharp peak at $x_{min}$, density at $x_{max}$ nearly vanishes. 
To understand this structure, let us consider a vicinity of a fixed point, for definiteness the fixed point $x_{min}$. 
Let us denote $\tilde x=x-x_{min}$. Then the first branch is $g_1(\tilde x)=A\tilde x$, while the second branch $g_2(\tilde x)=(B-A)\lambda/(1-A)+B\tilde x$. 
Thus in the vicinity of $\tilde x=0$, the branch $g_2$ takes the iterate of a small $\tilde x$ far away from the vicinity of the fixed point, while the branch $g_1$ brings the iterate of a small $\tilde x$ even closer to zero according to the stable linear mapping $\tilde x\to A \tilde x$. 
Correspondingly, the Frobenius-Perron equation \eqref{eq:fp0} for the invariant density around the left fixed point reduces to
\[
P(\tilde x)=\frac{1}{2A}P\left(\frac{\tilde x}{A}\right)
 \]
since the second term on the r.h.s. of \eqref{eq:fp0} gives zero contribution near $\tilde x=0$. Seeking the solution in the form
 $P(\tilde x)=C\tilde x^{\gamma-1}$ we obtain
 \begin{equation}
 \gamma=-\frac{\ln 2}{\ln A}\;.
 \label{eq:sing}
 \end{equation}
The power $\gamma$ defines the singularity of the density $P(x)$. 
One can see that the critical
value of the mapping slope is $A=1/2$. If $A<1/2$, then $\gamma<1$ and the distribution 
has a diverging peak near $x_{min}$. If $A>1/2$, then $\gamma>1$, and the density vanishes at the fixed point. 
The same condition holds for the right
fixed point at $x=x_{max}$: here the density has a peak for $B<1/2$, otherwise the density vanishes. For the case 
depicted  in Fig.~\ref{fig:f1}, we have $A<1/2$ and $B>1/2$, and thus the density has a peak near $x_{min}$ and 
vanishes near $x_{max}$.  The second mapping branch $g_2(x)$ "transports" the boundary peak at $x_{min}$ 
to the position $g_2(x_{min})$, which is subsequently split into two more which are located at $g_1(g_2(x_{min}))$ 
and $g_2(g_2(x_{min}))$ and so on, so that an infinite set of peaks  appears (with progressively smaller 
amplitudes since at each new iteration every peak is split into two).  It is interesting to note that the condition 
for the absence of peaks on both ends of the interval is that both $A,B>1/2$, which coincides with the 
above-mentioned necessary condition for a continuous density $AB>1/4$.

\subsection{Moments and autocorrelations of the damage distribution}
\label{sec:spd}
It is remarkable that although the distribution of damages in just-born cells is fractal, its statistical characterization in terms
of moments is quite simple and can be computed analytically for arbitrarily large asymmetry $a$. This, of course, is due to the linearity of the basic equation~\eqref{eq:fp0}.

We can compute moments $M_k$ of the distribution of $P(x)$ or the initial damage $N(x,0)$ by multiplying Eq.(\ref{eq:fp0}) by $x^k$ and integrating. After straightforward algebra, we get for the first two moments
\begin{equation}
M_1=\langle x\rangle = {\lambda(A+B)\over 2-A-B}\;,
\qquad M_2=\langle x^2\rangle = {2(A^2+B^2)\lambda\langle x\rangle+(A^2+B^2)\lambda^2 \over 2-A^2-B^2} \;.
\label{eq:avy}
\end{equation}
The variance is
\begin{equation}
Var(x)=M_2-M_1^2=\frac{2\lambda^2 (A-B)^2}{(2-A^2-B^2)(2-A-B)^2}\;.
\label{eq:vary}
\end{equation}
 {Note that these averages over a distribution of damage in a population at a given time are different from the lineage distributions mentioned in Sec.~\ref{sec:chaos} although the two are related \cite{thomas2017making,nozoe2017inferring}.}

Using the linearity of the governing kinetic equation~\eqref{eq:fp0} we can also compute the normalized auto-correlation function of the damages in just-born cells of different generations in a single lineage,
\[
C(m)=\frac{\av{(x(n)-\av{x})(x(n+m)-\av{x}}}{\text{Var}(x)},
  \]
  where the argument is the integer generation number. Calculation of the $C(m+1)$ with relation between $x(n+m+1)$ and $x(n+m)$ given by \eqref{eq:linmap}, leads to a recursion
  \[
  C(m+1)=C(m)\frac{A+B}{2}\;.
  \]
  This yields exponentially decaying correlations
  \begin{equation}
  C(m)=\left(\frac{A+B}{2}\right)^m\;.
  \label{eq:dcf}
  \end{equation}
In fact,  the full continuous-time correlation function presented Fig \ref{fig:acf} can also be calculated 
analytically (see \ref{ap:cf}).

\subsection{Phase-averaged distribution of damage}

Because of the linear relation $P(x)=T^{-1}N(x,0)$, we can use the properly  normalized $P(x)$ (i.e. $\int P(x)\;dx=1$) in the
two-dimensional damage density that in this case is factorized:
\[
n(x,\phi)=2\ln 2e^{-\ln 2\;\phi}P(x)\;.
\]
Thus, the marginal distribution over the phases is exponential
\[
w(\phi)=2\ln 2 e^{-\ln 2\;\phi}\;.
\]
We can also compute  the phase-averaged distribution $W(D)$ from Eq.\eqref{eq:wd}, however the result depends on the explicit form of the function $D=F(x,\phi)$.

\paragraph{Linear damage growth.}
Substitution of $F(x,\phi)=x+\lambda\phi$ in \eqref{eq:wd}  yields the following expression for the distribution of the damage integrated over the phases
\begin{equation}
W(D)=\frac{2\ln 2}{\lambda}\int_{D-\lambda}^D dx\; P(x) \exp\left[-\ln 2 \frac{D-x}{\lambda}\right]\;.
\label{eq:int1}
\end{equation}
As seen in Fig.~\ref{fig:f1}, unlike $P(x)$, $W(D)$  is continuous.  Using this expression, one can explicitly calculate the moments of the distribution damages:
\begin{equation}
\av{D}=M_1+\frac{\lambda}{\ln 2}(1-\ln 2),\qquad \text{Var}(D) =\text{Var}(x)+\left (\frac{\lambda}{\ln 2}\right)^2 (1-2(\ln 2)^2)\;.
\label{eq:dmom}
\end{equation}

\paragraph{Exponential damage growth.}
For $F(x,\phi)=x\exp[\beta T\phi]+\gamma\beta^{-1}(e^{\beta T \phi}-1)$
we get the following expression for the phase-integrated distribution of the damage
\begin{gather*}
W(D)=2\ln 2 \int dx\;d\phi\; 2\ln 2\;P(x) e^{-(\ln 2)\phi}\gamma(D-xe^{\beta T\phi}-\gamma\beta^{-1}(e^{\beta T \phi}-1))\\
=\frac{2\ln 2}{T(\gamma+\beta D)^{1+\frac{\ln 2}{\beta T}}}
\int_{x_{0}(D)}^D dx\; P(x) (\gamma+\beta x)^{\frac{\ln 2}{\beta T}},\quad 
x_0=De^{-\beta T}-\frac{\gamma}{\beta}(1-e^{-\beta T})\;.
\end{gather*}
With $\beta\to 0$ and $\lambda=T\gamma$ this reduces to the expression \eqref{eq:int1}. 

\section{Linear damage accumulation, damage-dependent lifetimes}

Here, we consider the same linear model with the mother-daughter damage inheritance relations \eqref{eq:linmap}, but assume that cell lifetimes depend on the damage, $\tau(x)$. 

\subsection{Stationary damage distribution}
We again start with the general Frobenius-Perron equation \eqref{eq:bc2} and rewrite it for the linear damage distribution model  \eqref{eq:linmap}: 
\begin{equation}
P(x)=\int dx'\; [\delta(x-Ax'-A\lambda)+\delta(x-Bx'-B\lambda)] P(x') e^{-\Gamma \tau(x')}\;,
\label{eq:frobp}
\end{equation}
or
 \begin{equation}
 P(x)=\frac{1}{A}P\left(\frac{x}{A}-\lambda\right)e^{-\Gamma \tau\left(\frac{x}{A}-\lambda\right)}+
 \frac{1}{B}P\left(\frac{x}{B}-\lambda\right)e^{-\Gamma \tau\left(\frac{x}{B}-\lambda\right)}\;.
\label{eq:frobp1}
\end{equation}

Unlike the previous section, the value of $\Gamma$ here is unknown and needs to be determined from the conservation of the total probability.
Numerically, it can be implemented iteratively  in two different ways. In the first, we solve Eq.~\eqref{eq:frobp1} 
iteratively starting from an arbitrary initial distribution $P_0(x)$, but at each iteration, we find $P_k(x,\Gamma)$ for 
a set of values $\Gamma$. The normalization condition $\int P_k(x,\Gamma)\;dx=1$ can be considered as 
equation for $\Gamma$, the root of which is determined numerically. Thus, the proper normalization is ensured at 
each iteration. As a result of these iterations of the Frobenius-Perron operator \eqref{eq:frobp}, we obtain a 
sequence of densities and of values of $\Gamma$, both of which converge. The corresponding limit are the stationary 
density and the corresponding stationary death rate. The second method is to start with some initial guesses 
for $P(x)$ and $\Gamma$, compute new $P(x)$ using (\ref{eq:frobp}), then compute 
$S=\int P(x,\Gamma)\;dx$ (which is generally not 1) and update $\Gamma=\Gamma\cdot S$. Then 
compute $P(x)$ in the next iteration, and do it until both $P(x)$ and $\Gamma$  converge to their 
asymptotic values. The second method is more computationally efficient and precise, however its convergence  
generally is not guaranteed. 

The stationary distribution  in Fig.~\ref{fig:f1}b obtained by solving the FP equation agrees very 
well with direct numerical simulation shown in Fig.~\ref{fig:scatter}d.  Qualitatively, it is also fractal, however, 
since formally equation \eqref{eq:frobp} does not correspond to a classical IFS, we cannot rely on the 
corresponding mathematical theory and compute fractal dimensions of the invariant measure analytically.  
Nevertheless, one can evaluate the generalized dimensions numerically. We illustrate this in Fig.~\ref{fig:f1}d, 
where we show the spectrum of fractal dimensions of asymptotic damage distributions for several values of 
parameter $\mu$ (while other parameters remain fixed).

\subsection{Cumulant expansion of the damage distribution}
In this section we present an approximate analysis of the Frobenius-Perron equation \eqref{eq:frobp1}  for the case of the linear dependence of lifetimes on damage, $\tau(x)=T(1+\mu (x-x_0))$. To simplify the calculation, without loss of generality we assume that $x_0=\av{x}_{\mu=0}$, the average damage of new-born cells for a fixed lifetime (see expression \eqref{eq:avy}). 

The main idea of the analysis below is to explore cases of weak lifetime variability across the population. As the expression for $\tau(x)$ shows,
this occurs if $\mu(x-x_0)$ is small in the range of $x$ values of the whole population, i.e. when either the parameter $\mu$ is small, or the deviations $(x-x_0)$ are small. A good measure of these deviations is the variance \eqref{eq:vary} that is proportional to $(A-B)^2$. Thus, the asymptotic analysis presented in this Section is valid either for weak dependence of lifetimes on the damage (small $\mu$) or weak asymmetry of damage segregation (small $\e=(A-B)^2$). 
 
 Below we will only sketch the theory, see \ref{expansions} for a full derivation. The method is based on expanding the characteristic function 
 \[
 C(k)=\av{e^{ikx}}=\int dx\;e^{ikx}\;P(x)\;.
 \]
Using the Frobenius-Perron equation \eqref{eq:frobp}, we can easily write the equation for $C(k)$:
 \begin{equation}
C(k)=e^{-\G T(1-\mu x_0)}[e^{ikA\l}C(Ak+i\mu\G T)+e^{ikB\l}C(Bk+i\mu \G T)]\;.
\label{eq:chf1}
\end{equation}

Taking the logarithm of both sides and introducing the  cumulant-generating function $F(k)=\log C(k)$, we obtain the following equation
\begin{equation}
\begin{gathered}
F(k)=-\G T(1-\mu x_0)+\ln (2)+\frac{1}{2}(ik(A+B)\l+F(Ak+i\mu\G T)+F(Bk+i\mu\G T))+\\
+\ln\cosh[\frac{1}{2}(ik\l(A-B)+F(Ak+i\mu\G T)-F(Bk+i\mu \G T))]\;.
\end{gathered}
\label{eq:cgf}
\end{equation}
Next we substitute the general cumulant expansion
\begin{equation}
F(k)=\sum_{m=1}^\infty \frac{c_m i^m k^m}{m!}
\label{eq:cum}
\end{equation}
and arrive at
\begin{equation}
\begin{gathered}
\sum_{m=1}^\infty \frac{c_m i^m k^m}{m!}=\\
-\G T(1-\mu x_0)+\ln\!2+\frac{ik\l(A+B)}{2}+\sum_{m=1}^\infty c_m i^m\frac{(Ak+i\mu\Gamma T)^m+(Bk+i\mu\Gamma T)^m}{2m!}\\
+\ln\cosh \left(ik\frac{\l(A-B)}{2}+\sum_{m=1}^\infty c_m i^m\frac{(Ak+i\mu\Gamma T)^m-(Bk+i\mu\Gamma T)^m}{2m!}\right)\;.
\end{gathered}
\label{eq:cumulants}
\end{equation}
Equating terms at powers of $k$, we obtain a system of equations for the cumulants.

Let us first briefly discuss the case of constant lifetimes $\mu=0$. In this case, equations in the order $m$
contain only cumulants with indices $m'\leq m$. Thus, the cumulants can be calculated sequentially starting from $c_1$.
In fact, this procedure is equivalent to the ad hoc derivation of moments in Sec.~\ref{sec:spd}.
Unfortunately, this property is lost for $\mu\neq 0$. However, as we shall argue below, in many interesting cases one
can perform a truncation of the infinite system of equations for the cumulants. Below we use a three-cumulants truncation: setting all $c_m ,m>3$  in \eqref{eq:cumulants} to zero gives a system of four equations for unknown $\G,c_1,c_2,c_3$: 
 \begin{align}
 \G T&=\ln\!2+\mu\G T(x_0-c_1)+\frac{c_2\mu^2(\G T)^2}{2}-\frac{c_3\mu^3(\G T)^3}{6}\;,\label{eq:f2-1}\\
 (2-A-B)c_1&=(A+B)(\l-c_2\mu\G T+\frac{1}{2}c_3\mu^2(\G T)^2)\;,\label{eq:f2-2}\\
 (4-2A^2-2B^2)c_2&=(A^2+B^2)(-2c_3\mu\G T)+(A-B)^2(\l+c_1-c_2\mu\G T)^2\;,\label{eq:f2-3}\\
 (4-2A^3-2B^3)c_3&=3c_2(A-B)(A^2-B^2)(\l+c_1-c_2\mu\G T)\;.\label{eq:f2-4}
 \end{align}
 Inspection of these equations reveals that there are indeed two potentially small parameters, justifying the truncation:
\begin{enumerate}
\item Small non-isochronicity of lifetimes, i.e. small parameter $\mu$. Cumulants in this case do not need to be small.
As one can conclude from \eqref{eq:f2-1}, $\G T$ is in fact represented by a power series in $\mu$ where higher
orders in $\mu$ come with higher cumulants. Also the higher cumulants enter \eqref{eq:f2-2} 
multiplied with powers of $\mu$. This allows for calculating $\G$ approximately,  as a series in $\mu$.
\item Small higher cumulants. This occurs if the difference $|x_{max}-x_{min}|=
\lambda (A-B)/[(1-A)(1-B)]$ is small. For even cumulants $c_{2k}$ one can deduce an upper bound
$c_{2k}\leq [(A-B)\lambda/2(1-A)(1-B)]^{2k}$  (this upper bound is 
achieved for a distribution in the form of two equal $\delta$ peaks at the end points of that 
interval). One expects a similar or even a smaller bound for the odd cumulants. This is consistent
with Eqs \eqref{eq:f2-3},\eqref{eq:f2-4}, from which it follows that $c_2\sim (A-B)^2$ and $c_3\sim (A-B)^4$. 
In this case, even for a finite $\mu$, the cumulant expansion \eqref{eq:f2-1} can be used to find a 
good approximation of $\G$ as a quickly converging power series in $\e=(A-B)^2$. This case was also treated in \cite{min2021transport} by a direct moment expansion of their general transport equation. 
\end{enumerate}

\begin{figure}
\centering
\includegraphics[width=0.49\textwidth]{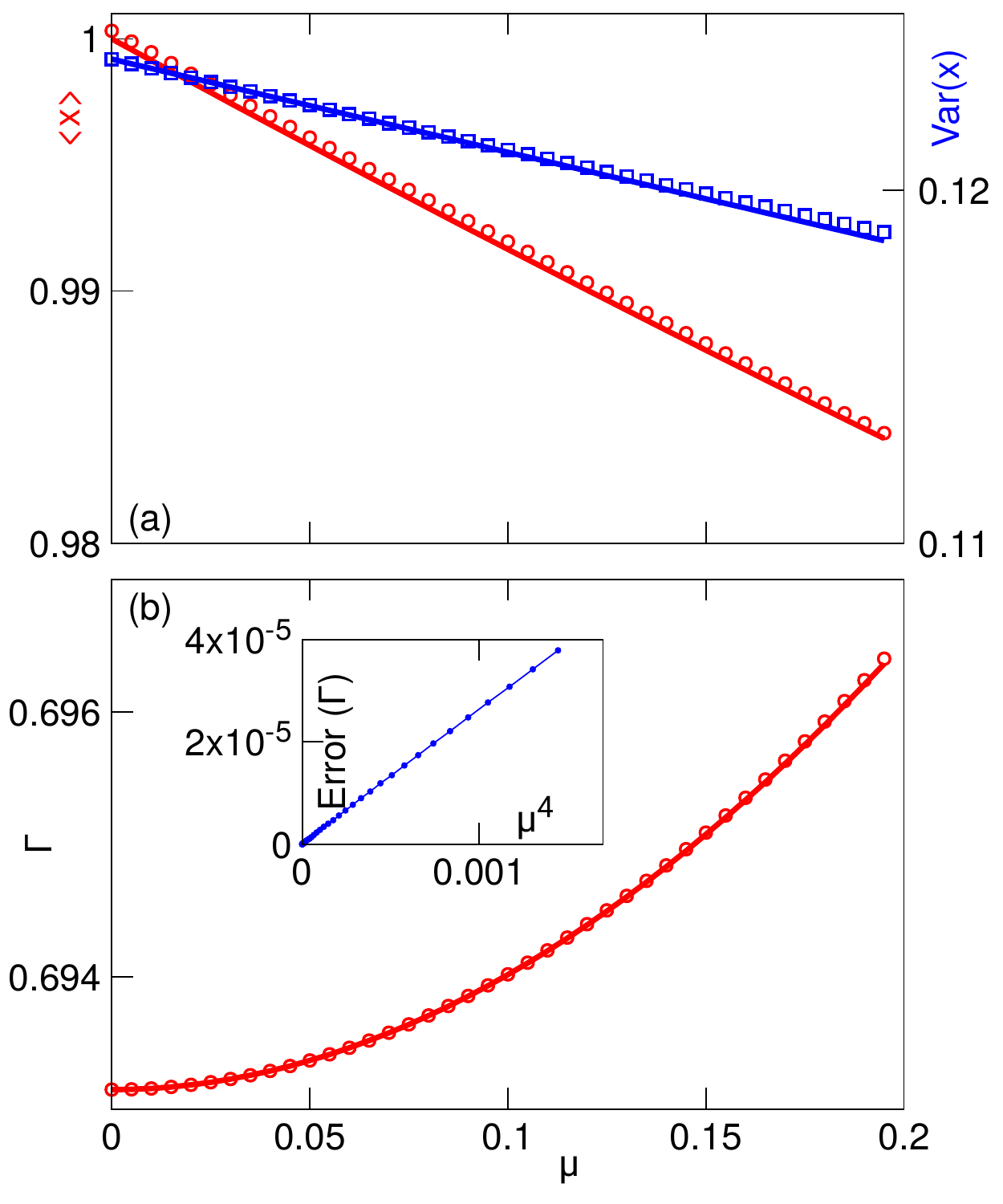}\hfill
\includegraphics[width=0.49\textwidth]{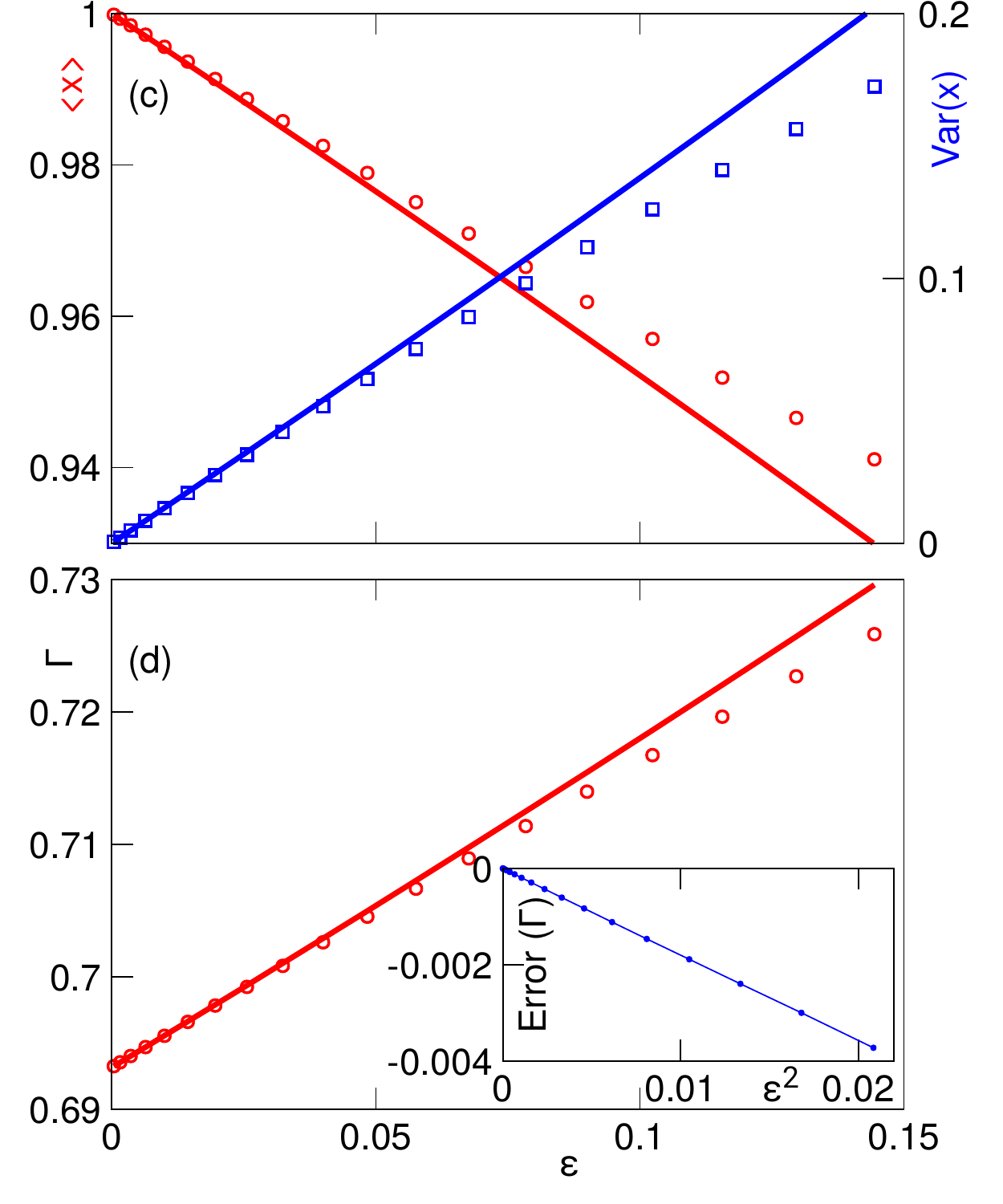}
\caption{Statistics of stationary damage distribution for small $\mu$ (panels (a,b)) or $\epsilon$
(panels (c,d)). 
Panel (a):  mean value $\av{x}$ (red circles) and variance $Var(x)$ 
(blue squares, right axis) of the stationary damage distribution in dependence on parameter $\mu$, 
together with the theory 
(solid lines). Parameters: $\l=T=1$, $A=0.35$, $B=0.65$. 
Panel (b): expansion of the population-averaged growth rate for small $\mu$. 
Red line: theory, red circles: numerical results. The inset shows difference between  
theory and numerics as function of $\mu^4$ (the same parameters as in panel (a)). 
Panel (c):  mean value $\av{x}$ (red circles) and variance $Var(x)$ 
(blue squares, right axis) of the stationary damage distribution in dependence on parameter $\e$, 
together with the theory (solid lines). Parameters: $\l=T=1$, $\mu=0.5$.
Panel (d): expansion of the population-averaged growth rate for small $\e$. The inset shows 
difference between  theory and numerics as function of $\e^2$ (the same parameters as in panel (c)).
}
\label{fig:muexp}
\end{figure}

For small $\mu$, we calculated the death rate $\G$ up to $O(\mu^3)$:
$\G T=\ln\!2+\mu^2  \G_2+\mu^3 \G_3$ (the term $\sim \mu$ is absent due to the proper choice of the
central value $x_0=\av{x}_{\mu=0}$ in the definition of $\tau(x)$). The explicit expressions for $\G_{2,3}$ are given in \ref{expansions}.  In Fig.~\ref{fig:muexp}a,b we compare this approximate analytical expression with numerics.

For small $\e$, the three-cumulants truncation only determine $\G$ up to $O(\e)$,
because $c_4\sim \e^2$ is missing in \eqref{eq:f2-1}. The calculated approximate expression
$\G T=\ln\!2+\G_1\e$ (see  \ref{expansions} for the explicit expression for $\G_1$) is compared with numerics in
Fig.~\ref{fig:muexp}c,d. 

The comparison presented in Fig.~\ref{fig:muexp} illustrates an excellent agreement between numerics and the asymptotic theory.  The results obtained here are also in agreement with earlier numerical and analytical findings \cite{chao2010model,chao2016asymmetrical,vedel2016asymmetric,min2021transport}. Specifically, while the mean damage $\langle x\rangle$ is a decreasing function of both $\mu$ and $\e$, the trends for the distribution variance are opposite. The increase in segregation asymmetry $\e$ obviously leads to a wider distribution of damages (greater variance). As we have seen in the previous section, for $\mu=0$ the mean damage of a population is independent of $\e$. A non-zero $\mu$ gives selective advantage to cells with less damage and thus reduces the mean for finite $\epsilon$ compared with $\e=0$.  The increase in $\mu$ for a fixed $\e$ gives greater selective advantage to cells with smaller damage and therefore reduces both the mean and the variance of the distribution. Also, the population-averaged growth rate $\Gamma$ is an increasing function of both $\mu$ and $\e$, that independently contribute to the spread among cells' lifetimes, at least when these two parameters are small, i.e. for a small spread of lifetimes in a population. Note that for more complex models relating cell growth and damage \cite{min2021transport}, the dependence of the population growth rate on the asymmetry may be non-monotonous and switch from beneficial to detrimental at some finite $\e$.  
 
\section{Nonlinear deterministic ADS  models}
Above we focused on models where many relevant statistical properties can be found analytically, either explicitly or as perturbative expressions for some small parameters. The main simplifying assumption is a linear redistribution of damages, which corresponds to linear Iterated Function Systems.
For more generic nonlinear models we do not have an analytical theory, and the goal of this section is to show
that numerical simulations reveal features similar to those in the analytically tractable cases.

\subsection{The Chao model}
As a representative example here we consider the Chao model~\cite{chao2010model}:
\begin{itemize}
\item \textbf{Damage gain}: The instantaneous damage in every cell grows 
linearly with time, $D(t;x)=x+\gamma (t-t_0)$  
\item \textbf{Lifetime} is the time when some product $P$ whose synthesis is suppressed by the damage according to 
$\dot P=1-s D(t;x)$ with $P(t_0)=0$, reaches a certain threshold value $P_0$.
One can easily see that the lifetime $\tau$ is a solution of the quadratic equation 
$(1-sx)\tau-\frac{s \gamma}{2}\tau^2=P_0$. Note, that this model of lifetime assumes that the product $sD$ remains small so that $\dot P>0$.
\item  \textbf{Damage inheritance}: Linear mapping $x_{1,2}=\frac{1\pm a}{2}y$ 
with constant $0<a<1$. 
\end{itemize}
Noteworthy, although the damage inheritance itself is linear, the effective map of damages of just-born cells is nonlinear because 
of the nontrivial lifetime dependence on damage.
Below, for compatibility with the theory above, we use $A=(1-a)/2$, $B=(1+a)/2$. 
The dependence $\tau(x)$ comes as a solution of the quadratic equation above
\begin{equation}
\tau(x)=\frac{1-s x}{s\gamma}-\sqrt{\left(\frac{1-s x}{s \gamma}\right)^2-\frac{2 P_0}{s \gamma}}\;.
\label{eq:chtim}
\end{equation}
Thus, the damage just prior to division is related to the initial damage as a nonlinear transformation
\[
y(x)=x+\gamma\tau(x)=\frac{1}{s}-\sqrt{(1/s-x)^2-2P_0\gamma/s}\;.
\]
and thus the two branches of the mapping of damage from one generation to the next are 
 \[
g_1(x)=Ay(x),\quad g_2(x)=By(x).
\]
Minimal and maximal values of $x$ are determined from $x_{min}=A y(x_{min})$ and $x_{max}=B y(x_{max})$:
\begin{gather*}
x_{min}=\frac{A}{s(1+A)}-\sqrt{\left(\frac{A}{s(1+A)}\right)^2-\frac{2P_0\gamma A^2}{s(1-A^2)}}\;,
\\
x_{max}=\frac{B}{s(1+B)}-\sqrt{\left(\frac{B}{s(1+B)}\right)^2-\frac{2P_0\gamma B^2}{s(1-B^2)}}\;.
\end{gather*}
\begin{figure}
\centering
\includegraphics[width=0.4\textwidth]{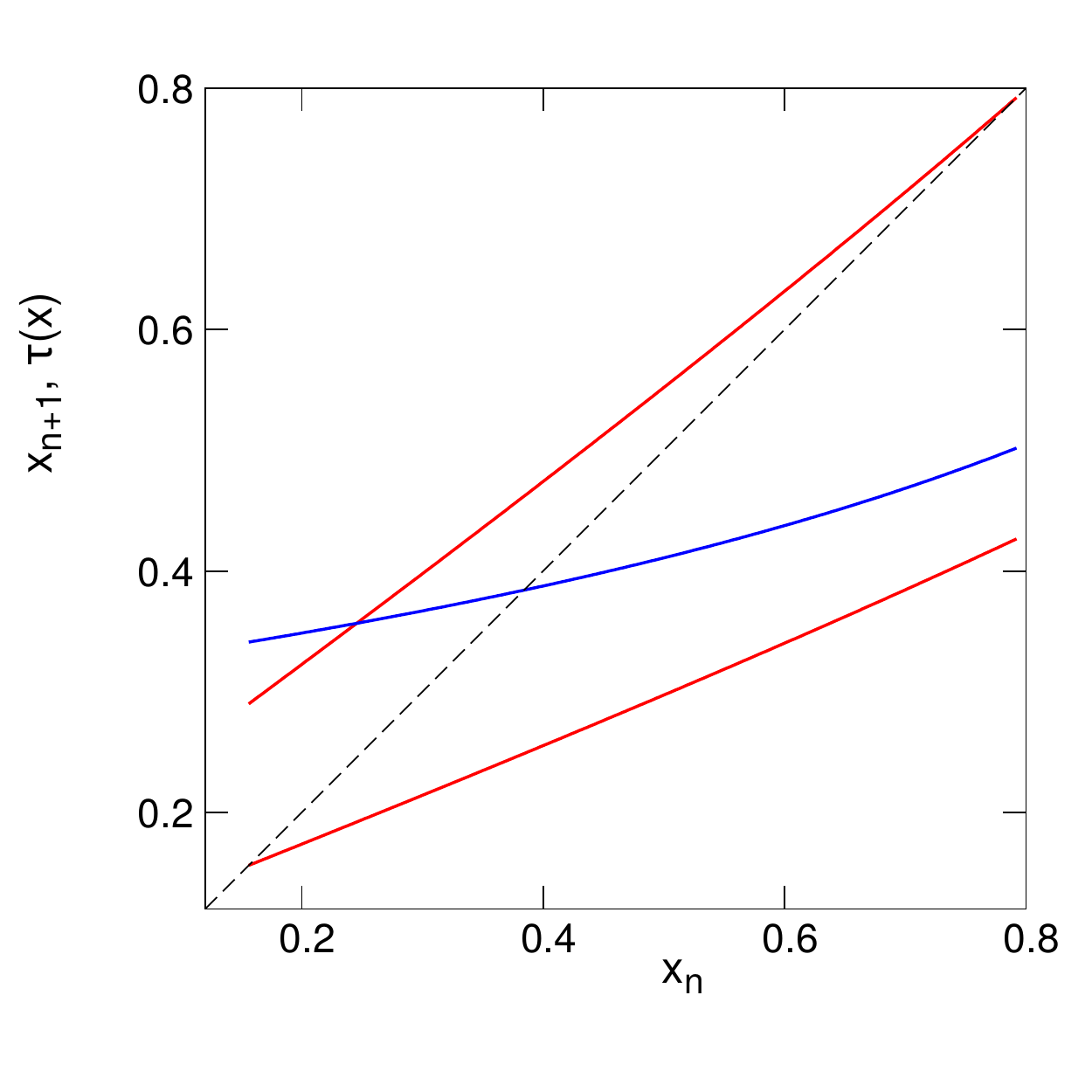}
\caption{Nonlinear mapping $g_{1,2}(x)$ (red lines) and $\tau(x)$ (blue line). Minimal and maximal
values of $x$ are determined by crossections of the red lines with the diagonal (dashed line).}
\label{fig:chmap}
\end{figure}

\begin{figure}
\centering
\includegraphics[width=0.7\textwidth]{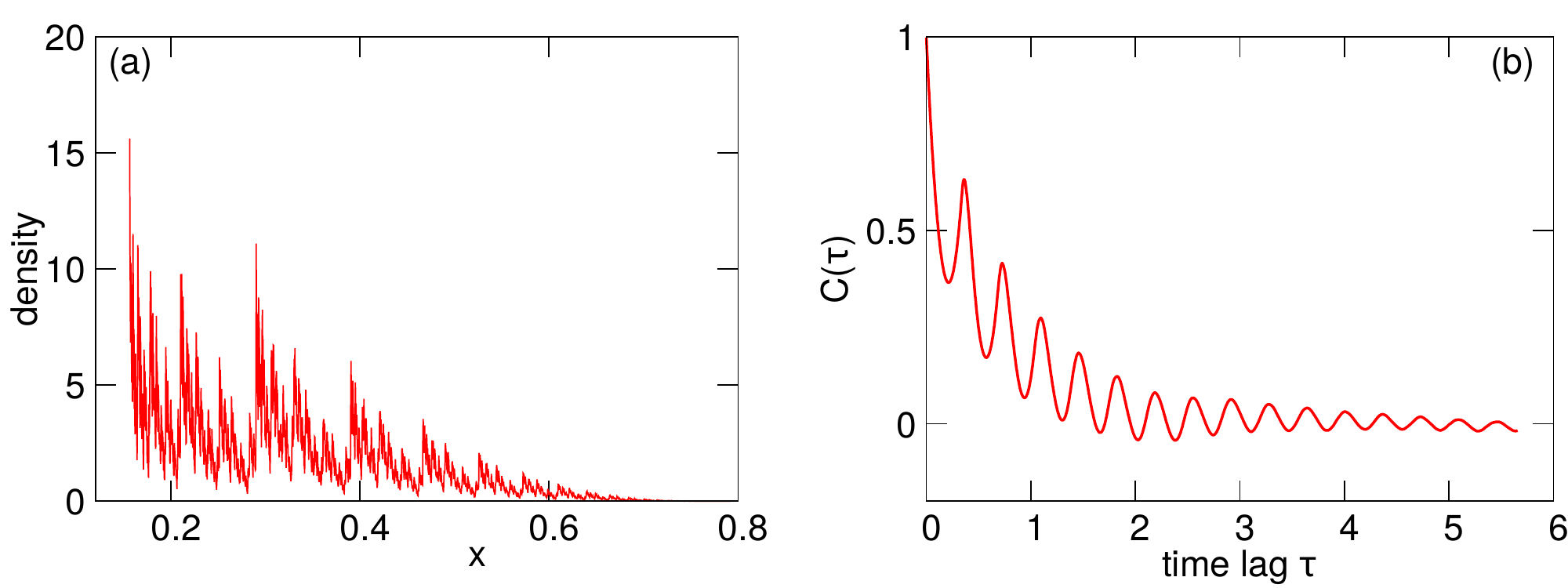}
\caption{Panel (a): Invariant density of $x$ for the map Fig \ref{fig:chmap}. Panel (b): the autocorrelation
function of a damage time series in a lineage.}
\label{fig:chdist}
\end{figure}
Figure \ref{fig:chmap} illustrates the transformation $g_{1,2}(x)$ for $A=0.35$, $B=0.65$, $s=0.4$, $P_0=0.3$, $\gamma=0.85$.
As before, it consists of two branches, but now the branches are nonlinear, furthermore, for the given parameters they overlap also along the vertical axis (this means that the mother's initial damage is not uniquely determined by the daughter's initial damage). Simulations of this model yield the distribution of $x$ shown 
in Fig. \ref{fig:chdist}a. As already discussed in Section \ref{sec:ifs}, maps like Fig.~\ref{fig:chmap} with a large overlap result in a peaky but formally
non-fractal (at least in the sense of absence of voids) distribution. However, the existing literature about fractal properties of IFS is mainly restricted to the linear case, thus exact statements about the fractal properties of the distribution are hardly possible. One can see from Fig. \ref{fig:chdist}a that the minima
of the distribution are separated from zero, which is an indication of a  density without voids. 
In this figure we also present the autocorrelation
function to confirm irregularity of the damage time dependence.

\subsection{General properties of nonlinear IFSs}
\label{sec:nifs}

\begin{figure}[!htb]
\centering
\includegraphics[width=\textwidth]{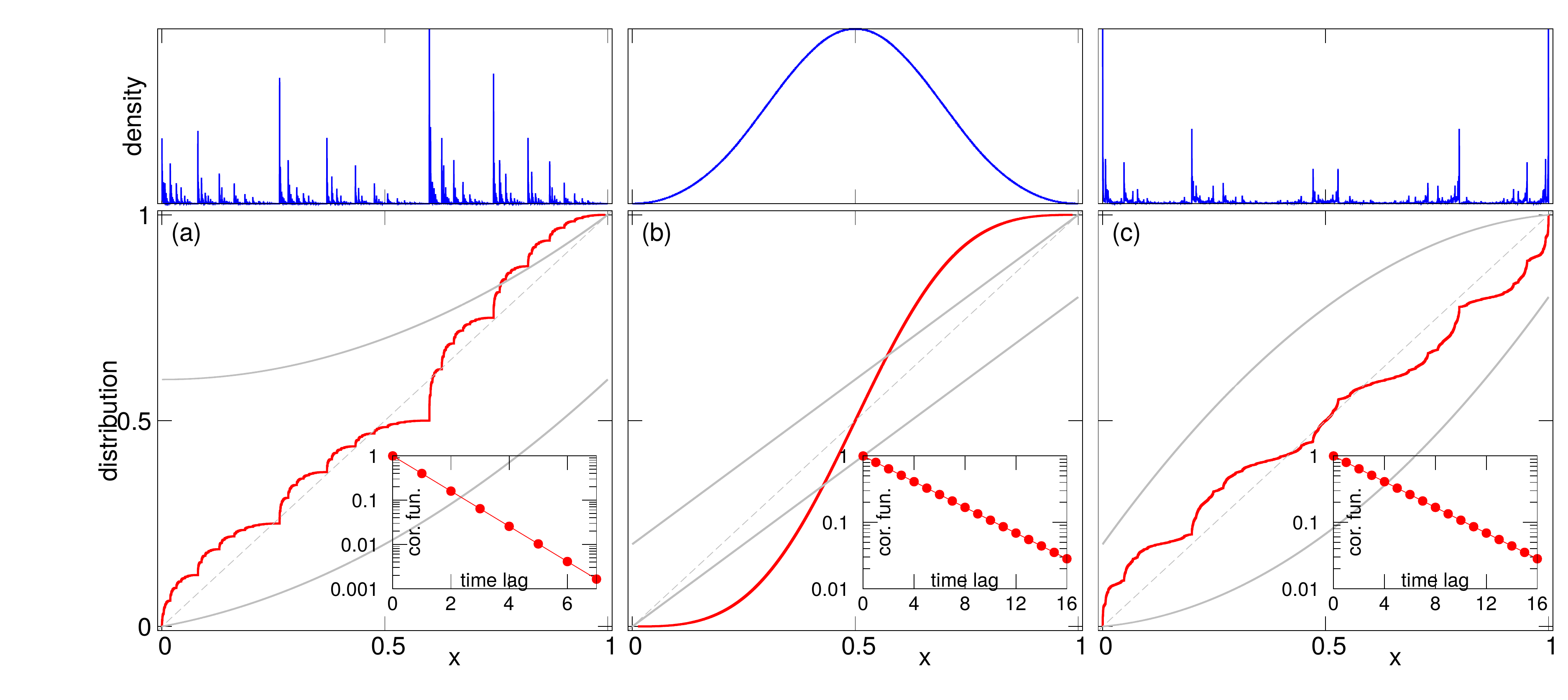}\\
\includegraphics[width=\textwidth]{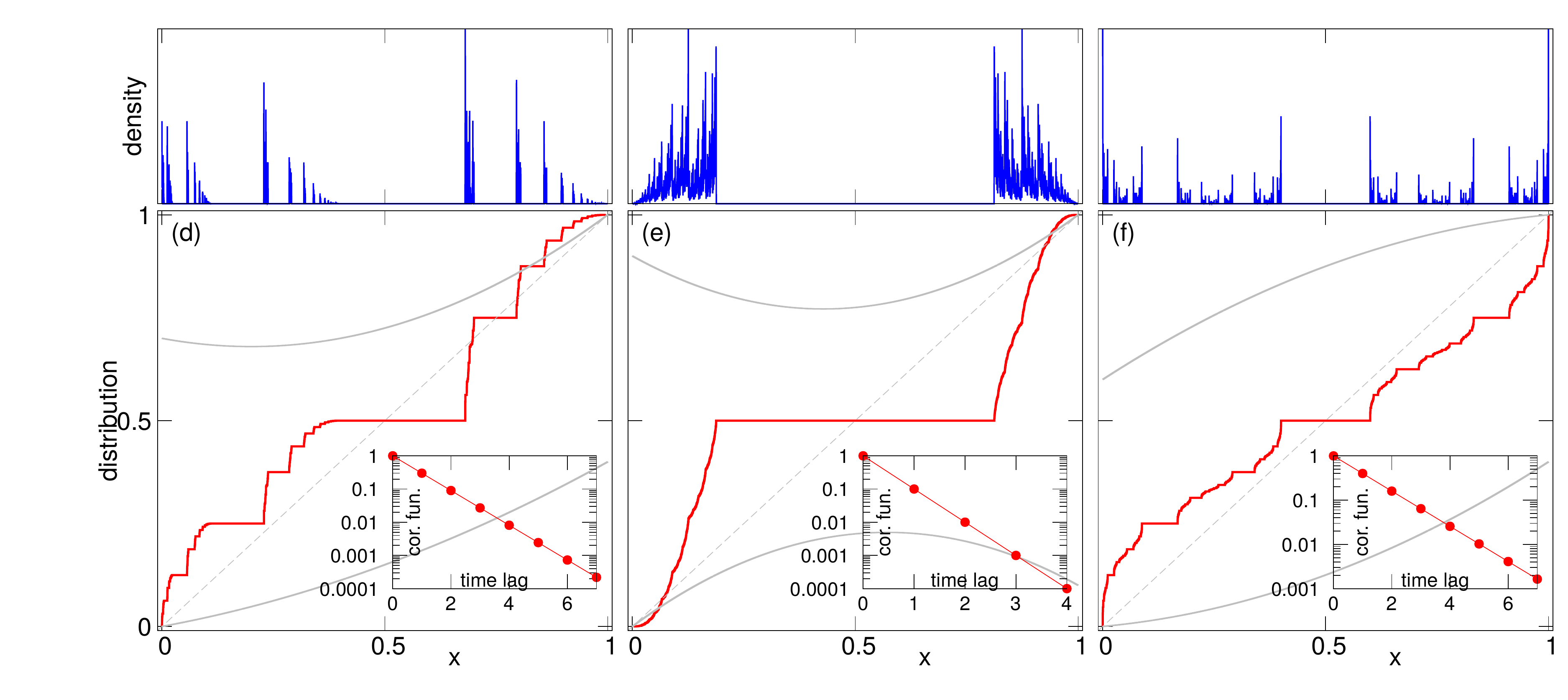}
\caption{Six examples of parabolic IFSs \eqref{eq:IFS}.  Panel (a): parameters $A=0.2$, $B=0.8$, $A_2=0.4$, $B_2=-0.4$. 
Panel (b): $A=B=0.8$,
$A_2=B_2=0.$ (in fact, this IFS is a linear one). Panel (c): $A=B=0.1$, $A_2=B_2=0.7$. 
Panel (d): $A=0.2$, $B=0.8$, $A_2=0.2$, $B_2=-0.5$.
Panel (e): $A=B=0.8$, $A_2=B_2=-0.7$. Panel (f): $A=B=0.1$, $A_2=B_2=0.3$. 
Red lines: the cumulative distributions $W(x)$. Blue lines: the densities $P(x)$. These densities
are depicted as histograms with $4096$ bins, thus the peaks are finite.
 The functions $f_{1,2}$ are depicted with grey lines; grey dashed line is diagonal. Insets show autocorrelation
 functions in linear-log presentation, to make evident the exponential decay of correlations. 
 The distributions are obtained from 65536 points starting from random initial conditions, 
 after transients of length 1000.} 
\label{fig:3ifs}
\end{figure}

Here we discuss some general statistical properties of distributions produced by nonlinear IFSs, without connecting them directly to particular
damage production and segregation models. 
IFS are characterized by two functions $g_{1,2}(x)$, so that in every iteration a given $x$ produces two new 
two states $x_{1,2}=g_{1,2}(x)$, thus the number of states $x$ in each generation $k$ grows exponentially as $2^k$.  We assume that $g_{1,2}(x)$ do not intersect, and $g_1(x)<g_2(x)$.  The interval of possible values of $x$
is limited by the two fixed point where $g_{1,2}(x)$ intersect the diagonal: $x_{min}=g_1(x_{min})$ and $x_{max}=g_2(x_{max})$. Without loss of generality we assume that $x$ is normalized such that $x_{min}=0$ and $x_{max}=1$, i.e. the distribution of $x$ lies within a unit interval $0\leq x\leq 1$. 

To illustrate the variety of distributions that can be generated by nonlinear IFSs, we consider parabolic functions $g_{1,2}(x)$: :
\begin{equation}
g_1(x)=Ax+A_2x^2,\quad g_2(x)=1-B(1-x)-B_2(1-x)^2
\label{eq:IFS}
\end{equation}
[here parameters $A,B$ define the slopes of $f_1(x)$ and $f_2(x)$ at the fixed points $x=0$ and $x=1$ and parameters $A_2, B_2$ characterize nonlinearity of the functions]. In Fig.~\ref{fig:3ifs} we present six typical regimes, for which we numerically computed the distribution densities $(P(x)$ and the corresponding cumulative distributions defined as $W(x)=\int_0^x P(x')dx'$. We also compute the autocorrelation functions of the iterative sequences of $x$.
In all cases the autocorrelation function~\footnote{Because lineages are not defined unambigously, in the calculations of the correlation function we used an ensemble average: 4096 points $x$ habe been created according the invariant distribution, and then
all decenders of these points in several generations were used for averaging.} decays exponentially, so that the damage level along a fixed lineage has strong 
chaotic properties. 

Fractal properties of the damage distribution strongly depend on the presence of an overlap or a gap between the ranges of
values of functions $g_{1,2}$ on the interval $0\leq x\leq 1$: these ranges are $0\leq g_1(x)\leq g_{1max}$ and
$g_{2min}\leq g_2(x)\leq 1$. 
Three situations without a gap are depicted in panels (a,b,c). The case of panel (a),  with a vanishing gap,
is qualitatively similar to the standard linear IFS with parameters $A+B=1$, discussed in Section~\ref{sec:ifs} above.
The measure is fractal, but it does not have voids. Panels (b,c) of Fig.~\ref{fig:3ifs} show two situations with an overlap,
they differ by the stability properties of the fixed points at $x_{min}$, $x_{max}$. Since near the fixed point nonlinear mappings can be linearized,  we can used the results of Section~\ref{sec:ifs}
(Eq.~\eqref{eq:sing}):
if $A>1/2$ and $B>1/2$ (case of panel (b)), the density at both  fixed points vanishes. This results in 
a rather smooth distribution which strongly resembles a Gaussian hump. On the contrary, for $A<1/2$, $B<1/2$ (case of panel (c)),
there are singularities at the fixed points, which are ``transported'' along the interval by functions $g_{1,2}$, so that the final
structure contains a sequence of peaks. We are not aware of any statements/conjectures about absolute continuity of the measure in such a nonlinear case.

Three situations with a gap $ g_{1max}<g_{2min}$ are depicted in panels (d-f). 
Cases (d,f) resemble a classical Cantor set with a large void in the center and  an hierarchy of smaller voids in the density distribution (these voids
correspond to the horizontal intervals of cumulative distributions in Fig.~\ref{fig:3ifs}). In panel (f) the fractal is symmetric (following the symmetry of the
functions $g_{1,2}$), while the distribution in panel (d) has a peak at $x_{min}$ and a vanishing density at $x_{max}$,
again according to the values $A,B$. In contradistinction to these cases, for non-monotonous functions $g_{1,2}$ in panel (e) we observe just one large void. The distribution density in this case is concentrated in two separated regions close to zero and close to one, and inside these regions
there are no additional voids (there the distribution is presumably continuous).

Returning to the asymmetric damage distribution problem, let us note that the borderline no gap/no overlap IFS of the linear model 1a2a3a mostly analyzed in the earlier sections is a degenerate situation caused by the assumption that the difference between the damages in two sister cells stays the same during the cells lifetimes. In general, even though during the cell division the damage is conserved, $f_2(y)=1-f_1(y)$, different damage gains in two sister cells (possibly caused by the autocatalytic nature of damage production or the lifetime dependence on the damage) lead to either a gap or an overlap between the two branches.   When the difference between the end damages in the sister cells  is smaller than their initial difference, ($y_2-y_1<x_2-x_1$), the corresponding IFSs have an overlap and typically produce a continuous, in some cases even rather smooth distributions.    If, on the other hand, the difference in damages becomes stronger over the lifetime ($y_2-y_1>x_2-x_1$) there is a gap 
 between the values of the two branches, and the distribution is typically fractal with voids (and even with a hierarchy of voids in the case of a Cantor-type measure). 

\section{Stochastic damage distribution and segregation}
\label{sec:stoc}
In biology, all processes are stochastic, due to extrinsic environmental fluctuations and intrinsic randomness of biochemical reactions that is particularly important on a sub-cellular scale. Randomness can in principle affect all three rules that constitute a model of asymmetric damage segregation. The fraction of the mother's final damage $y$ that a daughter cell inherits can be stochastic and governed by the conditional distribution $w_1(x|y)$. Note that since for a single mother cells there are two daughter cells, it is convenient to normalize this distribution as $\int w_1(x|y)dx=2$. Furthermore, because for two daughters' damages satisfy $x_1+x_2=y$, the conditional distribution should be symmetric $w_1(y-x|y)=w_1(x|y)$.

Damage gain and lifetime can also fluctuate and together they are specified by a two-dimensional distribution $w_2(y, \tau|x)$ conditioned on the initial damage $x$ (note that $y$ and $\tau$ are generally not independent, since longer-lived cells may on average accumulate more damage). However, it appears plausible to assume that segregation is statistically independent of the damage accumulation. Thus, the stochastic Frobenius-Perron equation for the probability distribution $P(x)$ can be written as follows
\begin{equation}
P(x)=\int dx'\;dy'\;d\tau'\;w_1(x|y')w_2(y',\tau'|x')e^{-\Gamma \tau'}P(x')\;,
\label{eq:frobps}
\end{equation}
where again the normalization condition $\int dx'\;P(x')=1$ yields the equation for the growth rate $\Gamma$. This equation is a more general form of the self-consistent equation for the distribution $P(x)$ derived in \cite{min2021transport} [they postulated a deterministic relationship between $x$ and $\tau$]. Note that if we take the probability distributions in the form 
\[
w_1(x|y')=\delta(x-f_1(y'))+\delta(x-f_2(y'));\quad w_2(y',\tau'|x')=\delta(y'-F(x'))\delta(\tau'-\tau(x'))\;,
\] 
we recover the  Frobenius-Perron equation \eqref{eq:bc2} for the fully deterministic case. 

\begin{figure}[h]
\centering
\includegraphics[width=0.33\textwidth]{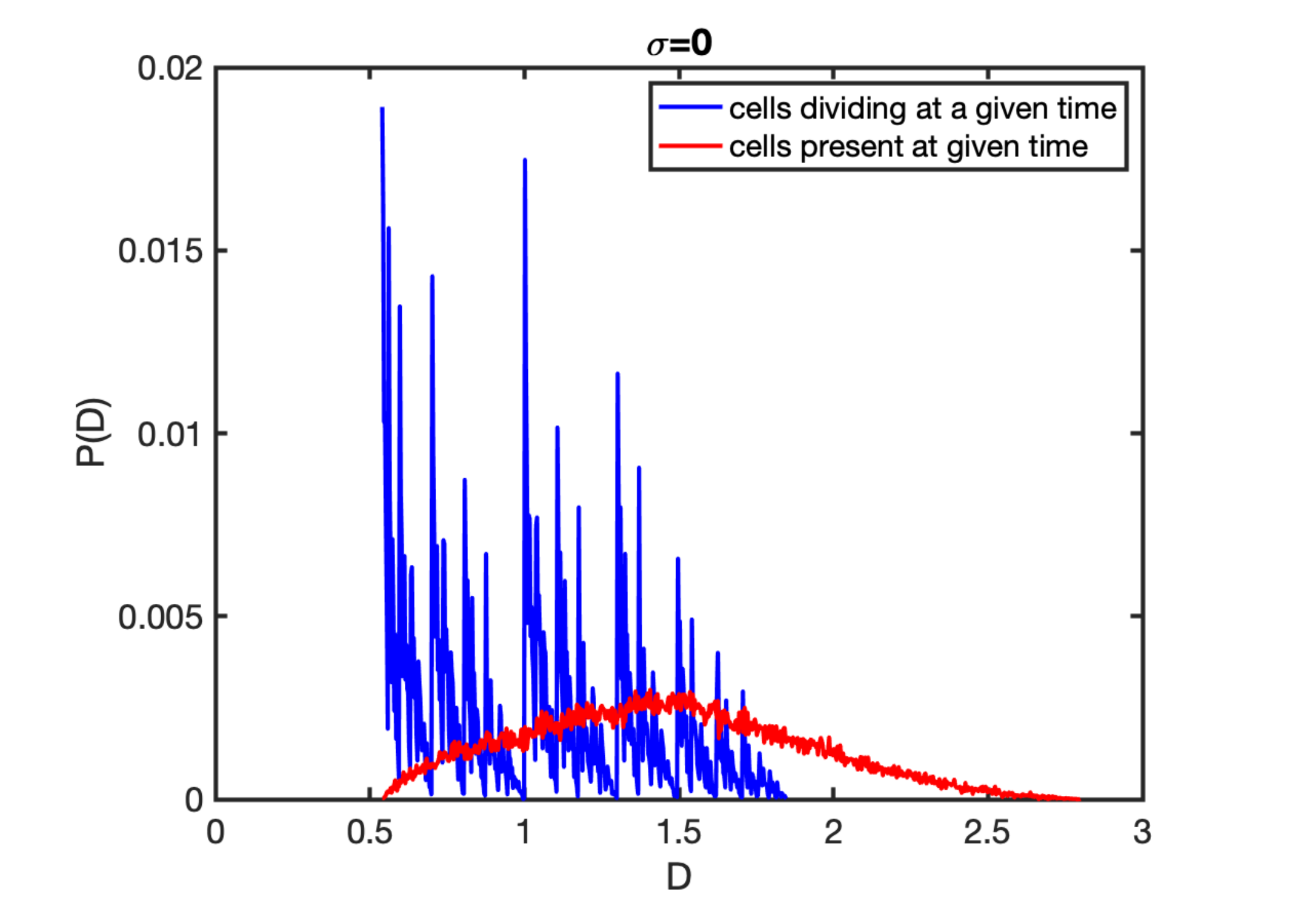}\hfill
\includegraphics[width=0.33\textwidth]{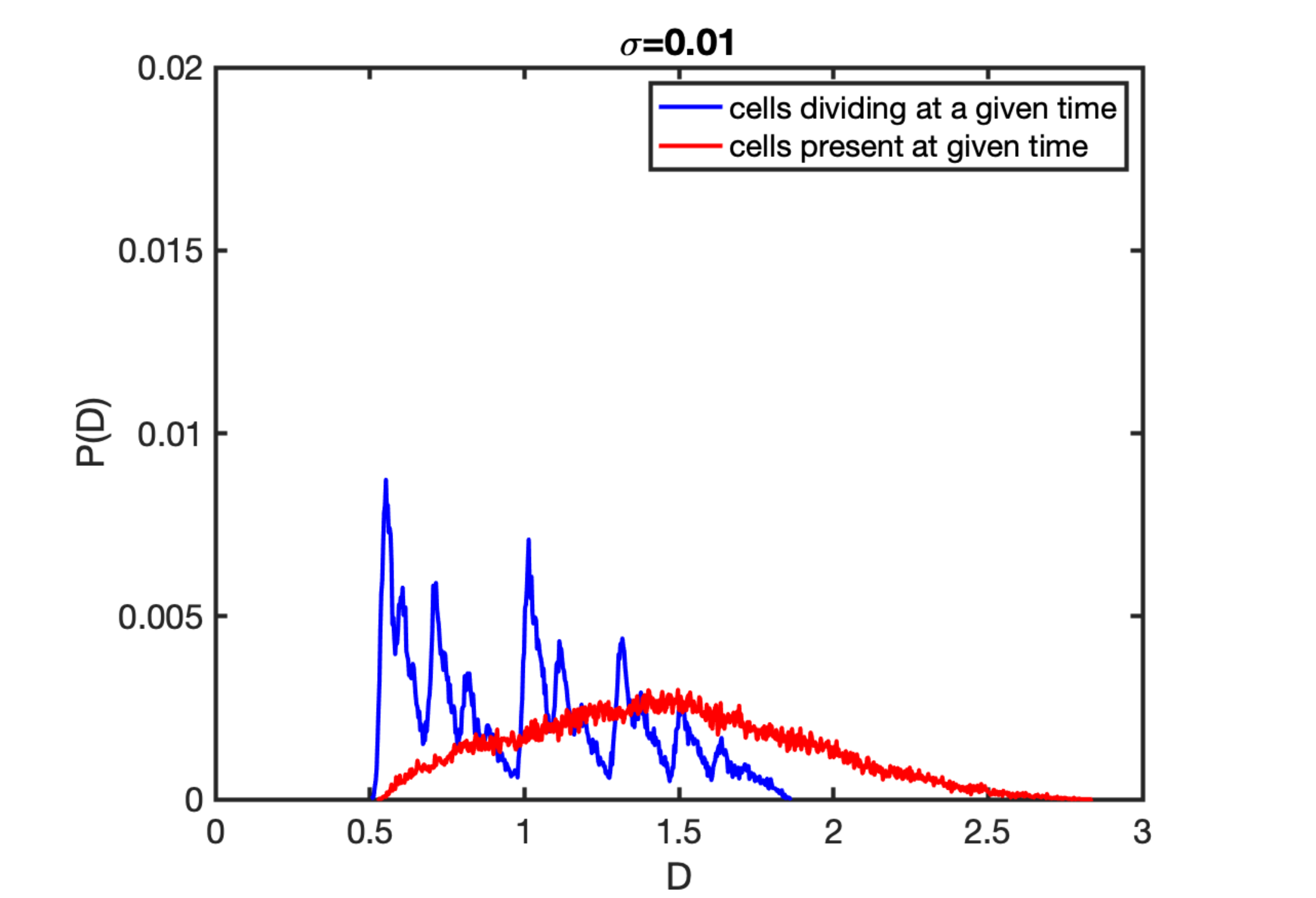}\hfill
\includegraphics[width=0.33\textwidth]{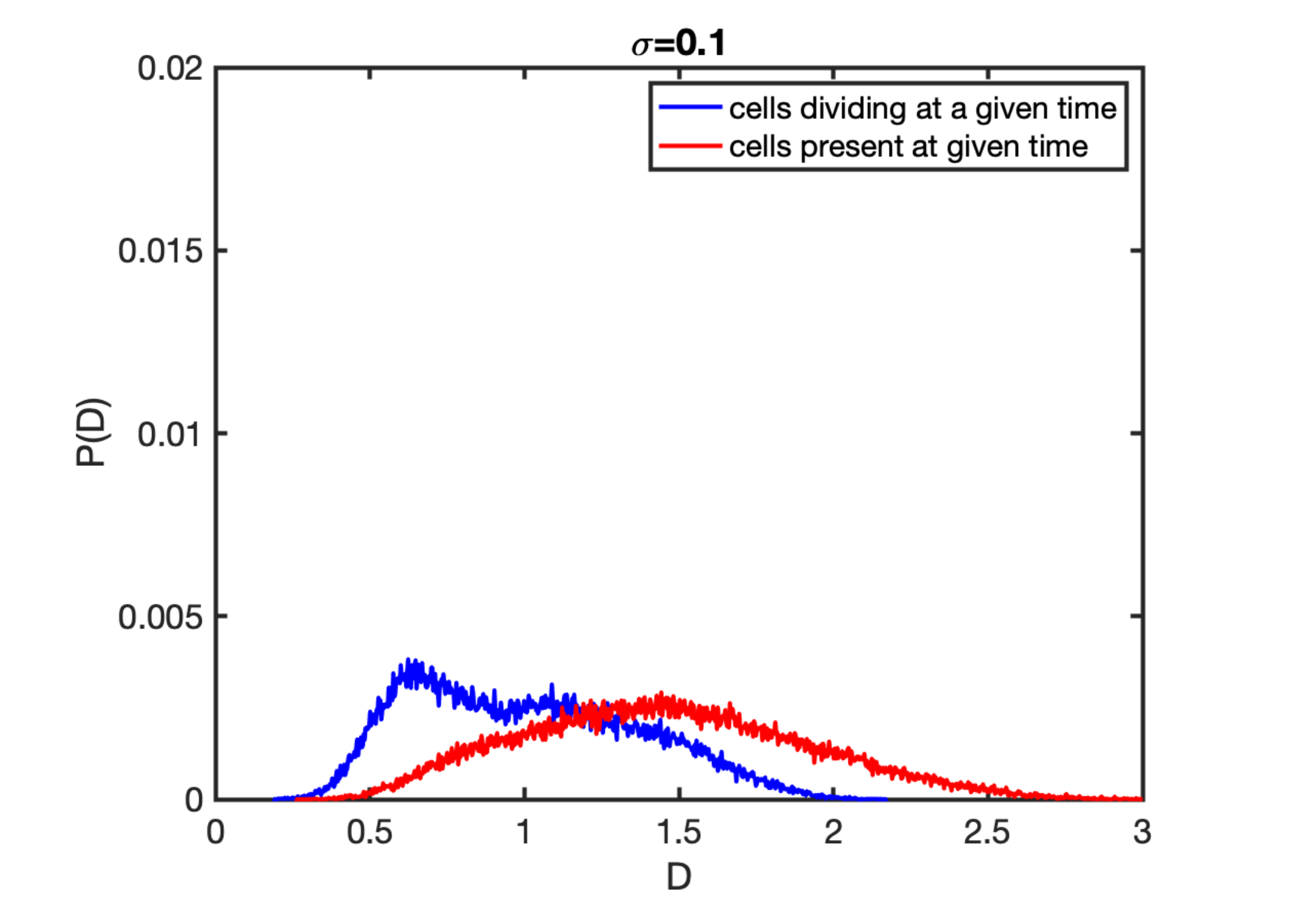}
\caption{ Initial damage distribution $P(x)$  obtained by simulating a Moran process for 1000 cells and $2\cdot 10^5$ divisions with deterministic linear damage accumulation $f_{1,2}(x)=(1\pm a)(x+\lambda)/2$ deterministic lifetime $\tau(x)=\tau_0(1+\mu x)$, and truncated Gaussian distribution \eqref{eq:Gauss_a} for the stochastic damage partition with $\lambda=1, a=0.3, \mu= 0.5, \tau_0=\log 2$
 and three values of $\sigma=0, 0.01, 0.1$.}
\label{fig:distr_gauss_a}
\end{figure}

If the stochasticity only affects the damage inheritance while the damage accumulation and the lifetime are deterministic functions of the initial damage, $y(x), \tau(x)$, we can substitute $w_2(y',\tau'|x')=\delta(y'-y(x'))\delta(\tau'-\tau(x'))$ in  \eqref{eq:frobps} and arrive at the Fredholm integral equation of the second kind
\begin{equation}
P(x)=\int dx'\;w_1(x|y(x'))e^{-\Gamma \tau(x')}P(x').
\label{eq:frobps1}
\end{equation}
that is equivalent to the self-consistent equation of \cite{min2021transport}. If the probability distribution $ w_1(x|y')$ is continuous but still close to the two-peaked form, for example
\begin{equation}
w_1(x|y')=\left\{\begin{array}{l} {w_0}[e^{-{(x-f_1(y'))^2\over 2\sigma^2}}+e^{-{(x-f_2(y'))^2\over 2\sigma^2}}],\quad 0<x<y'\;,\\
\\
					0,\quad \mbox{outside,}
						\end{array}
							\right.
	\label{eq:Gauss_a}
\end{equation} 
with small spread $\sigma$ ($w_0$ is the normalization constant to satisfy $\int w_1(x|y')dx'=2$), the damage distribution that was fractal in noise-free system, becomes continuous but still highly irregular (see  Fig.~\ref{fig:distr_gauss_a}). On the other hand, damage distributions that were continuous and smooth in deterministic limit, are much more robust agains noise (data not shown). Note that the distribution of damage in all (and not in just-born) cells at a given time is smooth even even when the distribution of initial damages is fractal, and so it is also very robust agains noise.  

The distribution of stochastic damage segregation can also be single-peaked. As an example, let us consider a truncated Gaussian
\begin{equation}
w_1(x|y')=\left\{\begin{array}{l} {w_0}e^{-{(x-y'/2)^2\over 2\sigma^2}},\quad 0<x<y\;,'\\
\\
					0,\quad \mbox{outside,}
						\end{array}
	\right.
	\label{eq:Gauss}
\end{equation}
and as before assume linear functions for the damage gain $y(x)=x+\lambda$ and lifetime $\tau(x)=\tau_0+\mu(x+\lambda)$. 
\begin{figure}[h]
\centering
\includegraphics[width=0.45\textwidth]{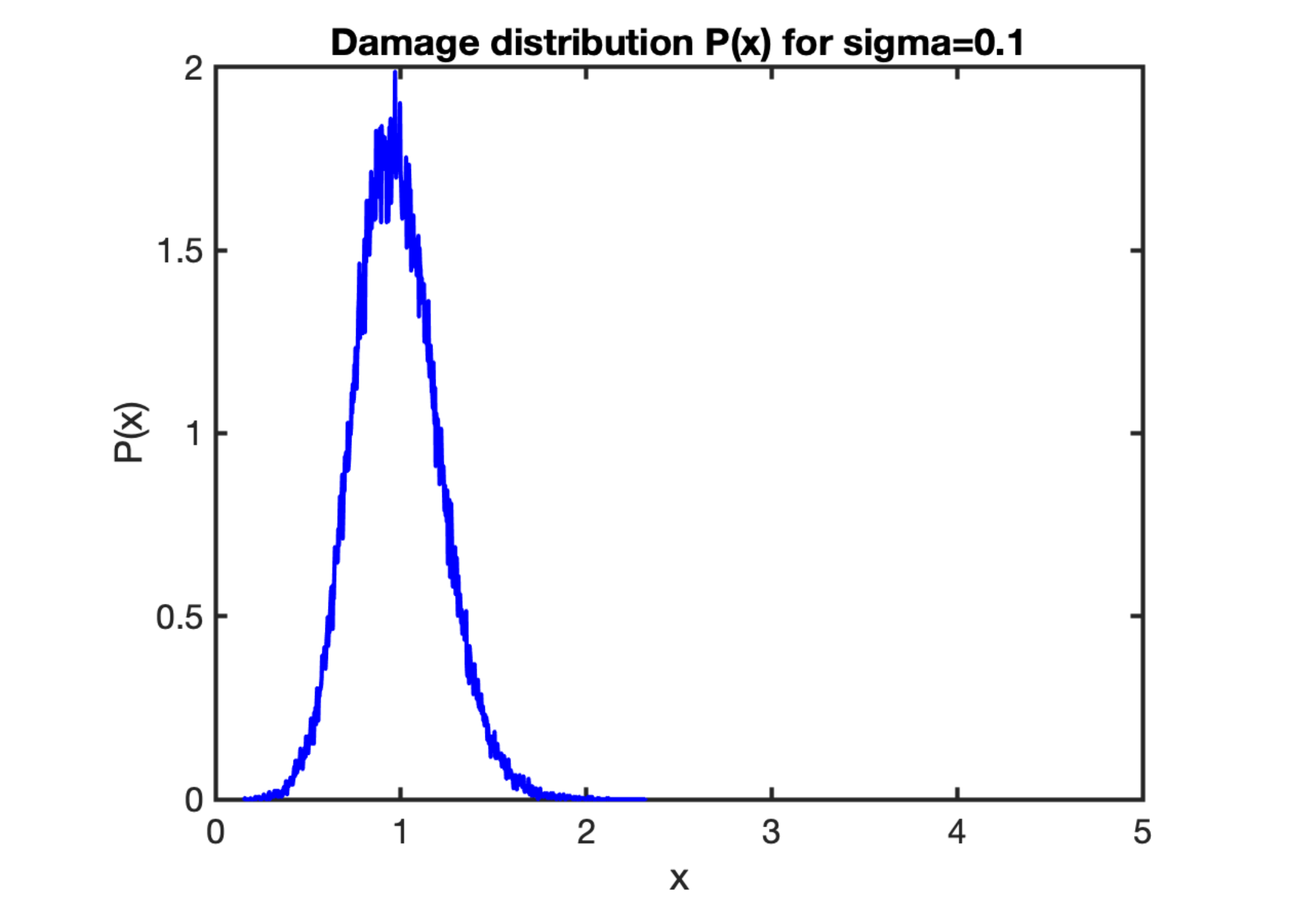}\hfill
\includegraphics[width=0.45\textwidth]{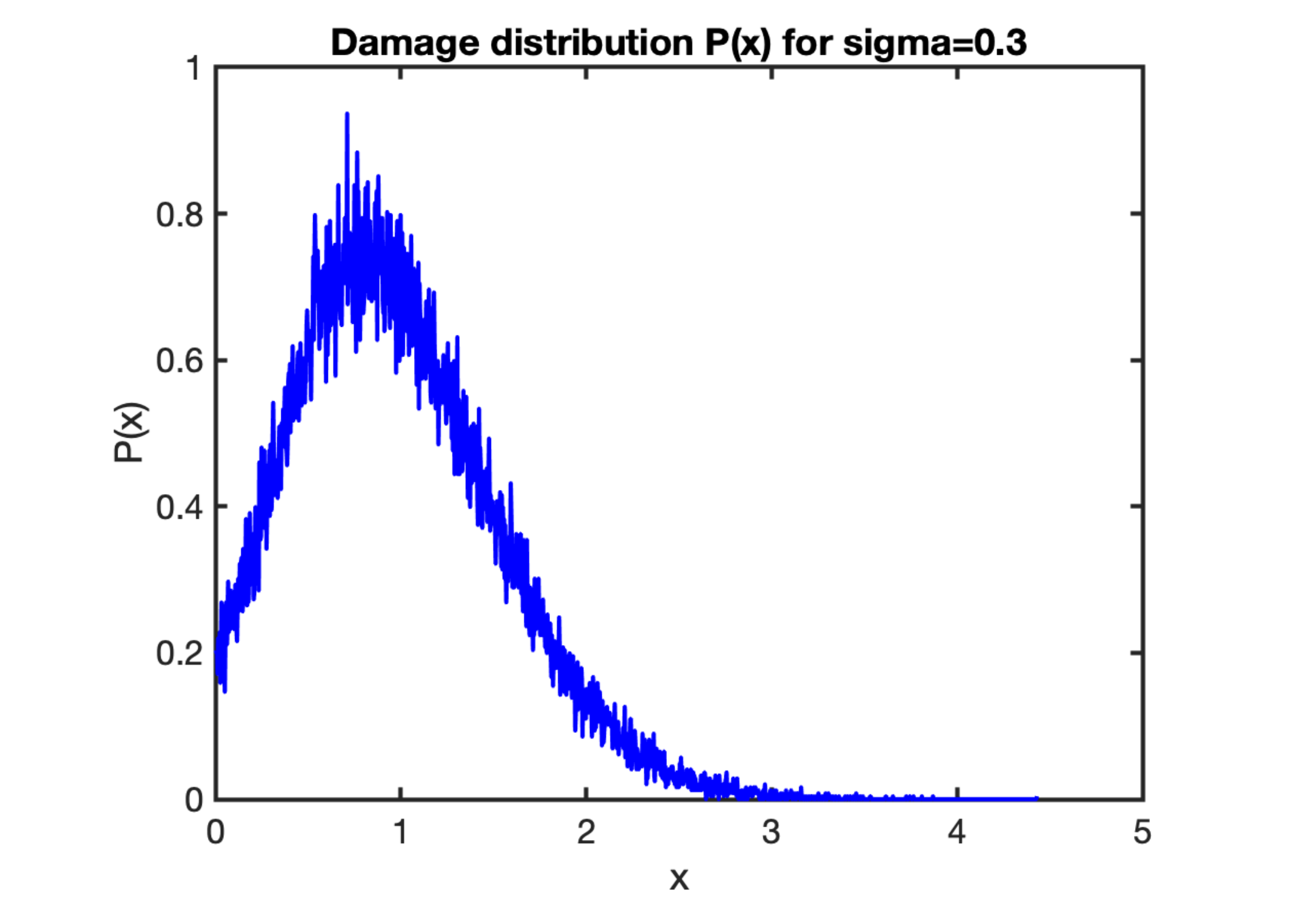}
\caption{ Initial damage distribution $P(x)$  obtained by simulating a Moran process for 1000 cells and $2\cdot 10^5$ divisions with deterministic linear damage accumulation and lifetime and truncated Gaussian distribution \eqref{eq:Gauss} for the stochastic damage partition with $\lambda=1, \mu= 0.5, \tau_0=\log 2$
 and $\sigma=0.1$ (left),  $\sigma=0.3$ (right).}
\label{fig:distr_gauss}
\end{figure}
\begin{figure}
\centering
\includegraphics[width=\textwidth]{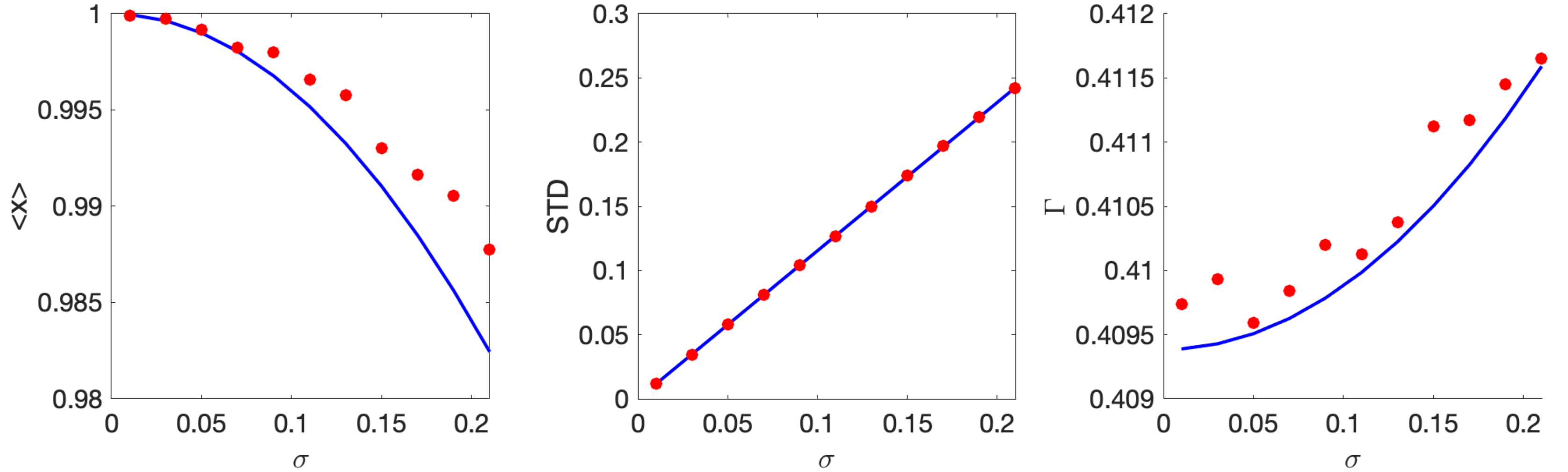}
\caption{Mean (left panel) and standard deviation (middle panel) of the initial damage distribution $P(x)$ and the corresponding population growth rate $\Gamma$ (right panel)
obtained by simulating a Moran process for 1000 cells and $2\cdot 10^5$ divisions with deterministic linear damage accumulation and lifetime and truncated Gaussian distribution \eqref{eq:Gauss} for the stochastic damage partition with $\lambda=1, \mu= 0.5, \tau_0=\log 2$. Blue lines are the theoretical values valid for small $\sigma$ according to formulas \eqref{eq:Gauss3}-\eqref{eq:Gauss5}.}
\label{fig:sigma_scan}
\end{figure}
Figure \ref{fig:distr_gauss} shows distributions of initial damage $P(x)$ for two different magnitudes of randomness $\sigma$. For small $\sigma$, the distribution $P(x)$ is narrow and appears close to a Gaussian.  In fact, it is easy to see that for small $\sigma$, when the truncation in \eqref{eq:Gauss} can be neglected, the solution of the FP equation \eqref{eq:frobps1} is Gaussian. Substituting 
\begin{equation}
P(x)={1\over \sqrt{2\pi}\Delta}e^{-{(x-x_0)^2\over 2\Delta^2}}
\label{eq:Gauss_sol}
\end{equation}
in \eqref{eq:frobps1} it is easy to solve for $x_0,\Delta$, and $\Gamma$ (see \ref{ap:stoc}). Fig. \ref{fig:sigma_scan} shows the mean and standard deviation of damage and the population growth rate as functions of $\sigma$ obtained from this approximate solution and directly from numerical simulations of the underlying stochastic ADS model. In agreement with earlier fundings \cite{chao2016asymmetrical,min2021transport}, the population-averaged growth rate $\Gamma$ increases with the randomness, which  should not be very surprising since inheritance randomness effectively creates asymmetry in damage separation. 

\begin{figure}
\centering
\includegraphics[width=\textwidth]{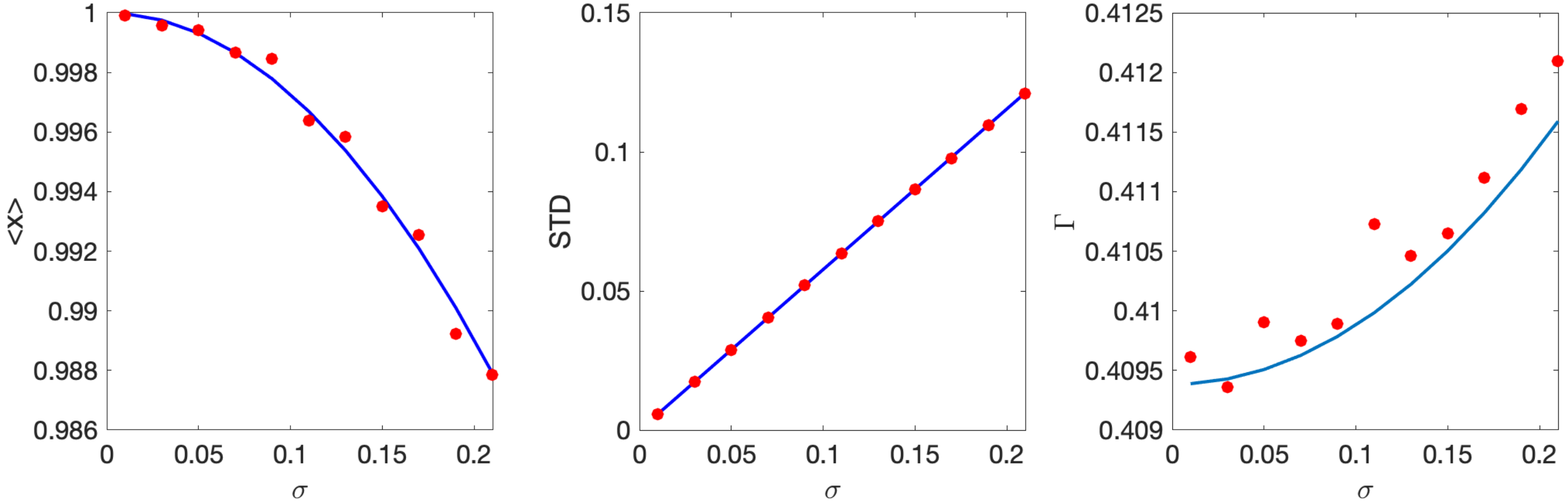}
\caption{Mean (left panel) and standard deviation (middle panel) of the initial damage distribution $P(x)$ and the corresponding population growth rate $\Gamma$ (right panel)
obtained by simulating a Moran process with deterministic symmetric damage partition, linear lifetime dependence on the final damage and  truncated Gaussian distribution \eqref{eq:Gauss_lambda} for the damage accumulation with $\lambda=1, \mu= 0.5, \tau_0=\log 2$  for 1000 cells and $2\cdot 10^5$ divisions. Blue lines are the theoretical values valid for small $\sigma$ according to formulas \eqref{eq:Gauss3lambda}-\eqref{eq:Gauss5lambda}.}
\label{fig:sigma_scan_lambda}
\end{figure}

In fact, asymmetry in cell division is not necessary for creating and exploiting the selective advantages of a broad damage distribution. Even if divisions are deterministic and symmetric, $w_1(x|y')=2\delta(x-y'/2)$, but the damage accumulation is stochastic, the damage distribution will have a finite width and the average growth rate will also be greater than in a purely deterministic symmetric case. In this case, the Frobenius-Perron equation \eqref{eq:frobps} simplifies to 
\begin{equation}
P(x)=\int dx'd\tau'\;w_2(2x,\tau'|x')e^{-\Gamma \tau'}P(x').
\label{eq:frobps2}
\end{equation}
Let us consider a simple illustrative example where the damage accumulation is described by a truncated Gaussian distribution and the lifetime $\tau$ is a deterministic linear function of the final damage $y$, $\tau=\tau_0+\mu y$:
\begin{equation}
w_2(y',\tau'|x')=\left\{\begin{array}{l} {w_0\over \sqrt{2\pi}\sigma}e^{-{(y'-x'-\lambda)^2\over 2\sigma^2}}\delta(\tau'-\tau_0-\mu y'),\quad y'>0,\tau'>0\;,\\
\\				0,\quad y'<0 \mbox{ or } \tau'<0\;.
						\end{array}
	\right.
	\label{eq:Gauss_lambda}
\end{equation}
For small $\sigma$ we again can ignore the truncation and solve the Frobenius-Perron equation \eqref{eq:frobps2} analytically by substituting the solution in the same Gaussian form \eqref{eq:Gauss_sol} (see  \ref{ap:stoc}).  The resulting $x_0,\Delta$, and $\Gamma$ as functions of $\sigma$ are shown in Fig.~\ref{fig:sigma_scan_lambda} superimposed  with the results of direct numerical simulations of the underlying stochastic process.

\section{Discussion}

Aging of microbial populations has been a subject of active research in recent years \cite{moger2019microbial}. While superficially fissioning bacterial cells appear immortal and divide symmetrically via binary fission, more close inspection reveals a slight phenotypic asymmetry that has been attributed to the asymmetric inheritance of damaged and aggregated proteins accumulated in the mother cell, among its daughters \cite{lindner2008asymmetric}. Importantly, a daughter cell inheriting a greater fraction of ancestor's damage, replicates slower than its sibling that inherits a smaller fraction of the damage. Cells that have a long line of ancestors inheriting lesser fraction of the damage, become ``rejuvenated'' and divide more often thus producing more offspring 
than ``age'' cells that have predominantly ancestors with greater fractions of damage accumulate larger amounts of damage. Asymmetry in damage inheritance has also been found in yeast \cite{aguilaniu2003asymmetric,coelho2014fusion} and even in higher eukaryotes \cite{rujano2006polarised}.  A number of conceptually simple models have been proposed to describe this branching process \cite{evans2007damage,erjavec2008selective,chao2010model,strandkvist2014asymmetric,vedel2016asymmetric,lin2019optimal,blitvic2020aging,min2021transport,levien2021non}. Simulations and analysis of these models  revealed that asymmetric damage segregation might be evolutionally preferred because it accelerates the mean population growth by letting rejuvenated cells, however most of these results were found from direct simulations. 

We re-examined these models focusing on understanding the mathematical and 
statistical properties of the damage distribution in populations of asymmetrically dividing microbes. 
 In a very broad class of models that encompasses both deterministic and stochastic ADS rules, the asymptotic damage distribution in the beginning of cell cycle can be described by a Frobenius-Perron-type equation \eqref{eq:frobps} where the rules of damage accumulation and inheritance are encoded in the transition probabilities    $w_1(x|y), w_2(y,\tau|x)$ for damage inheritance and damage accumulation, respectively. For deterministic transition rules, the damage distributions are broad and highly irregular resembling a fractal set. The mappings of the initial damage from generation to generation are equivalent to  to the Iterated Function Systems,  and for linear mappings the fractal dimensions of the stationary damage distributions can be computed analytically.  Stochasticity in damage accumulation and segregation smoothes out the fine fractal structure of the distributions, however for small noise, the distributions remain highly irregular, and their moments as well as the average growth rates of the population remain robust. 
  {We expect that experiments with fluorescent labeling of damaged proteins will reveal a complex multi-peaked structure of their distributions predicted by our theory. The overall width of the distribution will give us a quantitative measure of the asymmetry in damage inheritance, while the presence and magnitude of the distribution peaks (fractality) will characterize the degree of deterministic vs. stochastic asymmetry in damage distribution.}  {Another characteristics which
 is potentially measurable in experiments and has attracted less attention in previous literature, is the
 autocorrelation function of lineages \eqref{eq:corf}. Its structure sheds light on mixing properties of the ADS and can be compared with theoretical expressions,}
 
  {Here we would like briefly discuss the applicability of our stationary solutions for finite-size and growing populations. For a theoretical treatment, it is convenient to balance exponential population growth with cell death so the population reaches a stationary state in  a statistical sense thus allowing for the customary characterization of
stationary statistical processes (invariant distribution density, correlation function, Lyapunov exponent, etc.). It is easy to see that population-normalized damage distributions in exponentially growing and stationary populations are described by identical Frobenius-Perron (or transport) equations with death rate $\Gamma$ playing the role of mean growth rate $\Lambda$ (see \cite{min2021transport}). In finite-size population starting from a single cell (like in Fig.~\ref{fig:scatter} (a, b)) or a small group of cells, additional dependencies on the initial state and on the system size appear. We did not address them in the present study, but these transient aspects (which appear to be relevant for experimental observations)  definitely deserve future study.}
  
Phenotypic variability in clonal populations is often invoked as a useful bet-hedging strategy that improves the population survival chances in adverse environmental conditions \cite{veening2008bistability}.  {Phenotypic variability comes in many forms and can be caused by many factors such as mutations, multistability in underlying gene networks, noise in gene expression, or post-transcriptional processes, etc.  While not any phenotypic variability is beneficial,  the asymmetric damage segregation that maintains low damage in a fraction of the population to offset continuous accumulation of damaged proteins in all cells has been shown to help colony survival in stressful conditions when cells with sufficiently high of damage become mortal \cite{knorre2018replicative,proenca2019cell}.}

\section*{Acknowledgements}
Authors wish to thank Lin Chao and Terry Hwa for many useful discussions. LT has received partial financial support from the National Institute of Aging (grant R01 AG056440).

\section*{References}
\bibliographystyle{elsarticle-num}
\bibliography{ADS}
\appendix

\section{Analytical expression for the autocorrelation function of damage lineage}
\label{ap:cf}
Here we present an analytical expression for the autocorrelation function of the damage time series (Fig.~\ref{fig:traj}),
for the model with fixed lifetimes (Section~\ref{sec:flt}). Only for this model we have an analytic formula for the discrete
correlations of damages of just-born cells (Eq.~\eqref{eq:dcf}).
The  process $D(t)$  (Fig.~\ref{fig:traj}) can be represented as piecewise-linear function of time
\[
D(t)= x_n+\l \frac{t-nT}{T},\qquad nT\leq t \leq (n+1)T \;.
\]
Here $\av{x}=\l(A+B)(2-A-B)^{-1}$ (Eq.~\eqref{eq:avy}).
Let us represent it as a sum $D(t)=D_1(t)+D_2(t)+\av{x}+\l/2$ where
\[
D_1(t)=-\l/2+\l \frac{t-nT}{T},\qquad nT\leq t \leq (n+1)T 
\]
is purely periodic
and 
\[
D_2(t) = z_n,\qquad nT\leq t \leq (n+1)T 
\]
is purely random. Here $z_n=x_n-\av{x}$.
Each of processes $D_{1,2}(t)$ has zero mean.

The total correlation function $K(t)=\av{D(t')D(t'+t)}=\lim_{T\to\infty}T^{-1}\int_0^T D(t') D(t'+t) d t'$ 
is the sum of correlation functions. For the periodic part, we have to calculate the integral over the period
\begin{gather*}
K_1(t)=\int_0^{T-t}(-\l/2+\l \tau /T)(-\l/2+\l(\tau+t)/T)d\tau+\\
\int_{T-t}^T (-\l/2+\l \tau /T)(-\l/2+\l+\l(\tau+t)/T)d\tau=\\
=\l^2\frac{6(t/T)^2-6(t/T)+1}{12}\;,\qquad 0\leq t \leq T\;,
\end{gather*}
and this pattern repeats periodically. For the second part we use Eq.~\eqref{eq:dcf} $\av{z_n z_{n+p}}=V \gamma^p$ with
$\gamma=(A+B)/2$. Now because the function $D_2(t)$ is piecewise-constant, 
the integrals are evaluated trivially
\begin{gather*}
K_2(t)= V\left[\gamma^k((k+1)-t/T)+\gamma^{k+1}(t/T-k) \right],\quad k=0,1,2,\ldots,\quad kT\leq t\leq (k+1)T \;.
\end{gather*}
The overall correlation function is $K(t)=K_1(t)+K_2(t)$.
In Fig.~\ref{fig:acf} the normalized autocorrelation function $C(t)=K(t)/K(0)$  is compared with direct simulations.

\section{Asymmetric damage distribution in a Moran process}
\label{ap:moran}
In a Moran process~\cite{moran1958random}, whenever a cell divides into two daughter cells in the end of its lifecycle (at $\phi=1$), one additional cell is removed from the population at random, which keeps the number of cells in the population exactly constant at all times. Thus,  the corresponding intra-cycle death rate $\Gamma$ while still independent of the  damage and phase, is not a constant, but is determined at each moment of time self-consistently, depending on the current birth (doubling) rate.
We can integrate the time-dependent kinetic (master) equation \eqref{eq:ke} over all damages and phases to obtain the time derivative for the
total number of particles \eqref{eq:nc}
\begin{gather*}
\frac{d\mathcal{N}}{dt}=\int dx\int d\phi\left[-\partial_\phi (\w(x)N(x,\phi,t))-\Gamma(t) N(x,\phi,t)\right]=\\
-\Gamma(t)\mathcal{N}+\int dx\; \w(x)[N(x,0,t)-N(x,1,t)]\;.
\end{gather*}
On the other hand, from the transformation condition \eqref{eq:bc} it follows that
\[
\int dx \w(x)N(x,0,t)=2\int dx\w(x)N(x,1,t)\;.
\]
Thus from the condition that the total number of particles is constant, we obtain
\[
\Gamma(t)=\frac{\int dx\w(x)N(x,1,t)}{\int dx\int d\phi N(x,\phi,t)}\;.
\]
Asymptotically, the time dependence of $N(x,\phi, t)$ and $\Gamma(t)$ drops out, and we get
\[
\Gamma=\frac{\int dx\w(x)N(x,1)}{\int dx\int d\phi N(x,\phi)}\;,
\]
for which the same Eq.\~eqref{eq:station} applies. Thus, the Moran process asymptotically can be described using the same formalism with constant death rate $\Gamma$. 

\section{Cumulant expansion for a small lifetime spread}
\label{expansions}
In this appendix we describe approximate solutions of the system of the equations \eqref{eq:f2-1}-\eqref{eq:f2-4}
in terms of expansions in small parameters $\mu$ and $\e$.
\subsection{Expansion in $\mu$}
Because in \eqref{eq:f2-1} we keep accuracy $\mu^3$, we need to find $c_3,c_2,c_1$ in orders $\mu^0,\mu,\mu^2$,
respectively.
We substitute the expansions
\begin{gather*}
\G T=\G_0+\mu\G_1+\mu^2\G_2+\mu^3\G_3\;,\quad
c_1=c_{10}+\mu c_{11}+\mu^2 c_{12}\;,\quad
c_2=c_{20}+\mu c_{21}\;,\quad
c_3=c_{30}
\end{gather*}
in \eqref{eq:f2-1}-\eqref{eq:f2-4}. We start with the last equation \eqref{eq:f2-4}:
\begin{gather*}
(4-2A^3-2B^3)c_{30}=3c_{20}(A-B)(A^2-B^2)(\l+c_{10})\quad \Rightarrow\\
c_{30}=\frac{3c_{20}(A-B)(A^2-B^2)(\l+c_{10})}{(4-2A^3-2B^3)}
\end{gather*}
Next,  equation \eqref{eq:f2-3} yields
\begin{gather*}
(4-2A^2-2B^2)(c_{20}+\mu c_{21})=\\=
(A^2+B^2)(-2c_{30}\mu\G_0)+(A-B)^2((\l+c_{10})^2+2(\l+c_{10})(c_{11}-c_{20}\G_0)\mu)\quad \Rightarrow\\
c_{20}=\frac{(A-B)^2(\l+c_{10})^2}{4-2A^2-2B^2}\;,\quad
c_{21}=\frac{(A^2+B^2)(-2c_{30}\G_0)+(A-B)^2 2(\l+c_{10})(c_{11}-c_{20}\G_0)}{4-2A^2-2B^2}\;.
\end{gather*}
Equation for $c_1$ \eqref{eq:f2-2} yields:
\begin{gather*}
(2-A-B)(c_{10}+\mu c_{11}+\mu^2 c_{12})=(A+B)(\l-(c_{20}+\mu c_{21})\mu(\G_0+\mu\G_1)+
\frac{1}{2}c_{30}\mu^2\G_0^2)\quad \Rightarrow\\
c_{10}=\frac{\l(A+B)}{2-A-B}\;,\quad
c_{11}=\frac{(A+B)(-c_{20}\G_0)}{2-A-B}\;,\quad
c_{12}=\frac{(A+B)(-c_{21}\G_0-c_{20}\G_1)+\frac{1}{2}c_{30}\G_0^2}{2-A-B}\;.
\end{gather*}
Finally, the equation for $\G$ \eqref{eq:f2-1} reads:
\begin{gather*}
\G_0+\mu\G_1+\mu^2\G_2+\mu^3\G_3=\ln\!2+\mu(\G_0+\mu\G_1+\mu^2\G_2)(y_0-c_{10}-c_{11}\mu-c_{12}\mu^2)+\\
+\frac{(c_{20}+\mu c_{21})\mu^2(\G_0+\mu\G_1)^2}{2}-\frac{c_{30}\mu^3\G_0^3}{6}\;.
\end{gather*}
Here we use the freedom to choose $c_{10}=\frac{\l(A+B)}{2-A-B}$ (this quantity is $\mu$-independent).
Then the equation for $\G$ reads
\begin{gather*}
\G_0+\mu\G_1+\mu^2\G_2+\mu^3\G_3=\ln\!2+\mu^2(-\G_0 c_{11})-\mu^3(\G_1 c_{11}+\G_0c_{12})+\\
+\frac{(c_{20}+\mu c_{21})\mu^2(\G_0+\mu\G_1)^2}{2}-\frac{c_{30}\mu^3\G_0^3}{6}\;.
\end{gather*}
From this expression it follows that $\G_1=0$, and
\begin{gather*}
\G_0=\ln\!2\;,\quad
\G_2=-c_{11}\G_0+\frac{1}{2}c_{20}\G_0^2\;,\quad 
\G_3=-c_{12}\G_0+\frac{1}{2}c_{21}\G_0^2-\frac{1}{6}c_{30}\G_0^3\;.
\end{gather*}
Because the cumulants $c_{11},c_{20},c_{12},c_{21},c_{30}$ depend only on $\G_0$, 
the last expressions are explicit.
Especially simple appears the expression for $\G_2$:
\[
\G_2=(\l \ln\!2)^2\frac{(2+A+B)(A-B)^2}{(2-A-B)^2(2-A^2-B^2)}
\]

\subsection{Expansion in $\e$}
We rewrite here for convenience the 3-cumulant equations \eqref{eq:f2-1}-\eqref{eq:f2-4} 
and explicitly introduce parameter $\e=(A-B)^2$
\begin{align*}
 \G T&=\ln\!2+\mu\G T(y_0-c_1)+\frac{c_2\mu^2(\G T)^2}{2}-\frac{c_3\mu^3(\G T)^3}{6}\;,\\
 (2-A-B)c_1&=(A+B)(\l-c_2\mu\G T+\frac{1}{2}c_3\mu^2(\G T)^2)\;,\\
 (4-2A^2-2B^2)c_2&=(A^2+B^2)(-2c_3\mu\G T)+\e(\l+c_1-c_2\mu(\G T))^2\;,\\
 (4-2A^3-2B^3)c_3&=3c_2\e(A+B)(\l+c_1-c_2\mu\G T)\;.
 \end{align*}
 One can see that $c_2\sim\e$ and $c_3\sim\e^2$. We expect that also $c_4\sim\e^2$. Thus,
 in the 3-cumulant approximation only terms $\sim\e$ can be calculated correctly in $\G$.
 
 Therefore, we neglect the 3-rd cumulant and obtain
 \begin{gather*}
 \G T=\ln\!2+\mu\G T(y_0-c_1)+\frac{c_2\mu^2(\G T)^2}{2}\;,\\
 (2-A-B)c_1=(A+B)(\l-c_2\mu\G T)\;,\\
 (4-2A^2-2B^2)c_2=\e(\l+c_1-c_2\mu\G T)^2\;.
 \end{gather*}
 The last equation in order $\e$ yields $c_2=\e\frac{2\l^2}{(2-A^2-B^2)(2-A-B)^2}$.
 Substituting this in the first two equations and expanding $\G T=\G_0+\e\G_1$, $c_1=c_{10}+\e c_{11}$ we obtain
\begin{gather*}
 \G_0+\e\G_1=\ln\!2+\mu(\G_0+\e\G_1)(y_0-c_{10}-\e c_{11})+\e\frac{\l^2\mu^2\G_0^2}{(2-A^2-B^2)(2-A-B)^2}\;,\\
 (2-A-B)c_{10}+(2-A-B)\e c_{11}=(A+B)\l-\e\frac{2\l^2}{(2-A^2-B^2)(2-A-B)^2}\mu\G_0\;.
 \end{gather*}
  The solution with $y_0=c_{10}$ is
  \begin{gather*}
  c_{11}=-\mu\G_0\frac{2\l^2(A+B)}{(2-A^2-B^2)(2-A-B)^3}\;,\quad
  \G_0=\ln\!2\;,\\
  \G_1=-c_{11}\mu\G_0+\frac{\l^2}{(2-A^2-B^2)(2-A-B)^2}\;.
  \end{gather*}

\section{Gaussian approximation for stochastic ADS}
\label{ap:stoc}
Substituting
 \begin{equation}
w_1(x|y')={2\over \sqrt{2\pi}\sigma}e^{-{(x-y'/2)^2\over 2\sigma^2}}	\label{eq:Gauss1}
\end{equation}
and 
\begin{equation}
P(x)={1\over \sqrt{2\pi}\Delta}e^{-{(x-x_0)^2\over 2\Delta^2}}
\label{eq:Gauss_sol1}
\end{equation}
together with $\tau(x')=\tau_0+\mu(x'+\lambda)$ in \eqref{eq:frobps1} and integrating, we obtain
 \begin{equation}
{1\over \sqrt{2\pi}\Delta}e^{-{(x-x_0)^2\over 2\Delta^2}}=	{2\sqrt{2/\pi}\over \sqrt{\Delta^2 + 4\sigma^2}}
e^{-4x^2 +4 (x_0+\lambda-\Gamma\mu\Delta^2)x + 4\Delta^2\Gamma^2\mu^2\sigma^2 - 8\Gamma\mu\sigma^2(\lambda+x_0) - 2\Gamma \tau_0(\Delta^2 +4\sigma^2) - (\lambda+x_0)^2\over 2(\Delta^2 + 4\sigma^2)}.
\label{eq:Gauss2-1}
\end{equation}
Equating $O(x^2)$ terms in the exponents  on both sides, we obtain $\Delta^2=(\Delta^2 + 4\sigma^2)/4$
or
 \begin{equation}
\Delta^2=4\sigma^2/3\;.
\label{eq:Gauss3}
\end{equation}
Substituting \eqref{eq:Gauss3} in \eqref{eq:Gauss2-1} and equating terms $O(x^1)$ in the exponents, we obtain $2x_0=\lambda + x_0-4\Gamma\mu\sigma^2/3$ or
 \begin{equation}
x_0=\lambda-4\Gamma\mu\sigma^2/3\;.
\label{eq:Gauss4}
\end{equation}
Finally, substituting both \eqref{eq:Gauss3} and \eqref{eq:Gauss4} in \eqref{eq:Gauss2-1}  and equating terms $O(x^0)$, we get
\[
2\Gamma(\sigma^2\Gamma\mu^2 - \lambda\mu - \tau_0/2)=-\log 2,
\]
or
 \begin{equation}
\Gamma={2\lambda\mu +\tau_0 - \sqrt{4\lambda^2\mu^2 + 4\lambda\mu\tau_0 + \tau_0^2-8\sigma^2\mu^2\log 2}\over 4\sigma^2\mu^2}.
\label{eq:Gauss5}
\end{equation}
These dependences are shown by blue solid lines in Fig.~\ref{fig:sigma_scan}.

Similarly, for the case of symmetric partition and Gaussian damage accumulation distribution, we substitute \eqref{eq:Gauss_lambda} in \eqref{eq:frobps2} and again assume $\sigma$ is small and ignore the truncation: 
\begin{equation}
P(x)={w_0\over  \sqrt{2\pi}\sigma} e^{-\Gamma(\tau_0+2\mu x)} \int dx' e^{-{(2x-x'-\lambda)^2\over 2\sigma^2}} P(x').
\label{eq:frobps2a}
\end{equation}
Substituting $P(x)$ from \eqref{eq:Gauss_sol1} 
\[
P(x)={1\over \sqrt{2\pi}\sigma_p}e^{-{(x-\langle x\rangle)^2\over 2\sigma_p^2}}
\]
and integrating, we obtain
 \begin{equation}
{e^{-{(x-x_0)^2\over 2\Delta^2}}\over \sqrt{2\pi}\Delta}=
{e^{-4x^2 + 4(x_0+\lambda-\Gamma\mu(\sigma^2+\Delta^2))x - 2(\sigma^2+\Delta^2)\Gamma \tau_0 - (\lambda+x_0)^2\over 2(\Delta^2 + \sigma^2)}\over \sqrt{2\pi}\sqrt{\Delta^2 + \sigma^2}}\;.
\label{eq:Gauss2-2}
\end{equation}
Equating $O(x^2)$ terms in the exponents, we obtain 
 \begin{equation}
\Delta^2=\sigma^2/3\;.
\label{eq:Gauss3lambda}
\end{equation}
Substituting \eqref{eq:Gauss3lambda} in \eqref{eq:Gauss2-2} and equating terms $O(x^1)$ in the exponents, we obtain 
 \begin{equation}
x_0=\lambda-4\Gamma\mu\sigma^2/3\;.
\label{eq:Gauss4lambda}
\end{equation}
Finally, substituting both \eqref{eq:Gauss3lambda}  and \eqref{eq:Gauss4lambda} in \eqref{eq:Gauss2-2}  equating terms $O(x^0)$, we get
 \begin{equation}
\Gamma={2\lambda\mu +\tau_0 - \sqrt{4\lambda^2\mu^2 + 4\lambda\mu\tau_0 + \tau_0^2-8\sigma^2\mu^2\log 2}\over 4\sigma^2\mu^2}\;.
\label{eq:Gauss5lambda}
\end{equation}
These dependences are shown by blue solid lines in Fig.~\ref{fig:sigma_scan_lambda}. Interestingly, formulas \eqref{eq:Gauss4lambda}, \eqref{eq:Gauss5lambda} for the mean 
initial dmage $x_0$ and the growth rate $\Gamma$ coincide 
with expressions \eqref{eq:Gauss},\eqref{eq:Gauss5} while 
the formula \eqref{eq:Gauss3lambda} for the width of the initial damage distribution $\Delta$ is different from \eqref{eq:Gauss3}.


\end{document}